\documentclass[final,5p,times,twocolumn]{elsarticle}

\usepackage{amsmath,amssymb,bm}
\usepackage{graphicx}
\usepackage{booktabs}
\usepackage{multirow}
\usepackage{microtype}
\usepackage[hidelinks]{hyperref}
\usepackage{xcolor}
\usepackage{array}
\usepackage{placeins}
\usepackage{siunitx}
\usepackage{xcolor}

\journal{Physics Letters B}

\newcommand{\Heff}{H_{\mathrm{eff}}}
\newcommand{\Htr}{H_{\mathrm{tr}}}
\newcommand{\Eq}{E_{\mathrm{q}}}
\newcommand{\Eref}{E_{\mathrm{ref}}}
\newcommand{\Ec}{E_{c}}
\newcommand{\MeV}{\,\mathrm{MeV}}
\newcommand{\keV}{\,\mathrm{keV}}

\newcommand{\ket}[1]{\lvert #1\rangle}
\newcommand{\FG}{\text{FG--R6}}

\begin{document}
\begin{frontmatter}

\title{Nuclear scattering phase shifts with factorized geometric-time RODEO
on a quantum processor}

\author[knu]{Myeong-Hwan Mun}
\author[ssu]{Jubin Park\corref{cor1}}
\ead{honolov@ssu.ac.kr}
\author[ssu]{Myung-Ki Cheoun\corref{cor2}}
\ead{cheoun@ssu.ac.kr}
\author[hu]{Eunja Ha}

\cortext[cor1]{Corresponding author}
\cortext[cor2]{Corresponding author}

\address[knu]{Department of Physics, Kyungpook National University, Daegu 41566, Republic of Korea}
\address[ssu]{Department of Physics and Origin of Matter and Evolution of Galaxies Institute, Soongsil University, Seoul 06978, Republic of Korea}
\address[hu]{Department of Physics and Research Institute for Natural Sciences,
Hanyang University, Seoul 04763, Republic of Korea}

\begin{abstract}
We reconstruct elastic $s$-wave neutron--proton ($np$) phase shifts from
trapped spectra measured on IBM Aachen using a compressed RODEO circuit.
For a schematic square-well interaction represented by classically
constructed four-dimensional effective Hamiltonians, four positive-energy
levels at five trap strengths supply twenty inputs to classical
modified effective range expansion (MERE) extrapolation to free space.
The six-cycle factorized geometric-time RODEO implementation (\FG)
combines ancilla reuse, 
exact query-phase separation, post-compilation binding, and
six numerically optimized geometric evolution times.
Numerical cycle-count tests support this choice within the adopted
local spectral tolerances.
At the same three-qubit width and equal shot budgets, \FG\ reduces
median compiled depth and two-qubit-gate count by $33.1\%$ and
$37.2\%$ relative to the dynamic direct RODEO implementation 
with ten cycles (direct R10), 
increases the median fitted
amplitude, and lowers median finite-shot energy uncertainty from
$0.610$ to $0.534\keV$.
For the present dataset and primary fit on the $0.1$--$30.0\MeV$ grid,
the maximum central phase-shift deviation from exact-energy MERE
decreases from $0.842^\circ$ to $0.163^\circ$, despite a larger
root-mean-square (RMS) trapped-energy deviation.
The full-grid gain arises mainly in the low-energy extrapolation region;
direct R10 has slightly smaller residuals relative to the same reference
on the $10.0$--$30.0\MeV$ common energy-interpolation subset.
The \FG\ central curve differs from the analytical square-well
solution by at most $0.156^\circ$.
Fit-form sensitivity remains appreciable and is assessed separately
from finite-shot uncertainty.
This reduced-space benchmark connects RODEO circuit compression to
nuclear continuum observables and shows why circuit performance must
be assessed through the scattering reconstruction and its energy range,
not spectral RMS errors alone.
\end{abstract}


\end{frontmatter}

\section{Introduction}
\label{sec:introduction}

Elastic-scattering phase shifts probe nuclear interactions in the
continuum.
A unified description of bound states and scattering requires nuclear
correlations and appropriate asymptotic boundary conditions, a continuing
challenge despite advances in \textit{ab initio} reaction theory
\cite{Johnson2020,Nollett2007,Quaglioni2008}.
Weak harmonic confinement allows bound-state methods to address
scattering by replacing the continuum with discrete, normalizable states
\cite{Luu2010,Zhang2020}.
For short-range interactions in an uncoupled elastic channel, the
modified effective range expansion (MERE) relates these trap-dependent
levels to free-space scattering, including finite-confinement corrections.
For $s$ waves, extrapolation to vanishing confinement yields the
effective-range function $p\cot\delta_0(E)$.
The eigenenergies of the confined system can be estimated using quantum eigensolvers, while the subsequent fitting in energy and extrapolation in confinement strength are performed classically \cite{Wang2024}.

The physical significance of the trapped-spectrum approach lies not 
in reproducing isolated eigenvalues, but in connecting 
quantum-computed discrete spectra to continuum observables 
\cite{Luu2010,Zhang2020,Wang2024}. Scattering phase shifts encode 
the interaction over a continuous energy interval and determine 
experimentally accessible quantities such as cross sections, resonance 
positions, and widths. Their reconstruction combines information from 
several trapped levels and confinement strengths, so spectral errors 
can be amplified, suppressed, or reorganized by the subsequent 
energy interpolation and zero-confinement extrapolation. Consequently, 
the performance of a quantum eigensolver cannot, in general, be 
assessed from level-by-level energy errors alone. Carrying the measured 
spectrum through the complete trapped-spectrum analysis therefore 
provides a more physically relevant benchmark of whether a 
resource-reduced quantum circuit preserves the information required 
for nuclear scattering.

The RODEO algorithm filters energy components through repeated
ancilla-controlled time evolution and postselection
\cite{Choi2021,Qian2024}.
Wang \textit{et al.}\ used it to extract phase shifts for model
neutron--proton ($np$) and neutron--alpha ($n\alpha$) scattering
from trapped spectra in noiseless simulations \cite{Wang2024}.
Complementary hardware studies employed variational eigensolvers and
quantum subspace expansion in finite oscillator spaces \cite{Sharma2024},
or real-time evolution \cite{Turro2024}.
The trapped-spectrum approach requires controlling both the cost of
estimating several positive-energy levels and the propagation of their
errors to the free-space phase shift.

Our recent benchmark of a single trapped $np$ level \cite{Mun2026} implemented the same ideal ten-cycle RODEO filter with twelve active qubits in a static circuit on IonQ Forte-1 and three in an ancilla-recycled dynamic circuit on IBM Aachen, reducing circuit width by $75.0\%$.
Mid-circuit measurement and reset recycle one ancilla \cite{DeCross2023},
making the auxiliary-qubit requirement independent of the cycle count
and leaving more qubits for the nuclear system, but retaining the
same controlled-evolution sequence.
The interleaved IBM scan reproduced the trapped energy and its
finite-confinement MERE input within finite-shot uncertainties,
although independent runs showed additional variation. 
That calculation supplied one finite-confinement input; a free-space
extraction requires several levels and trap strengths.

\begin{figure*}
  \centering
  \includegraphics[width=0.71\textwidth]{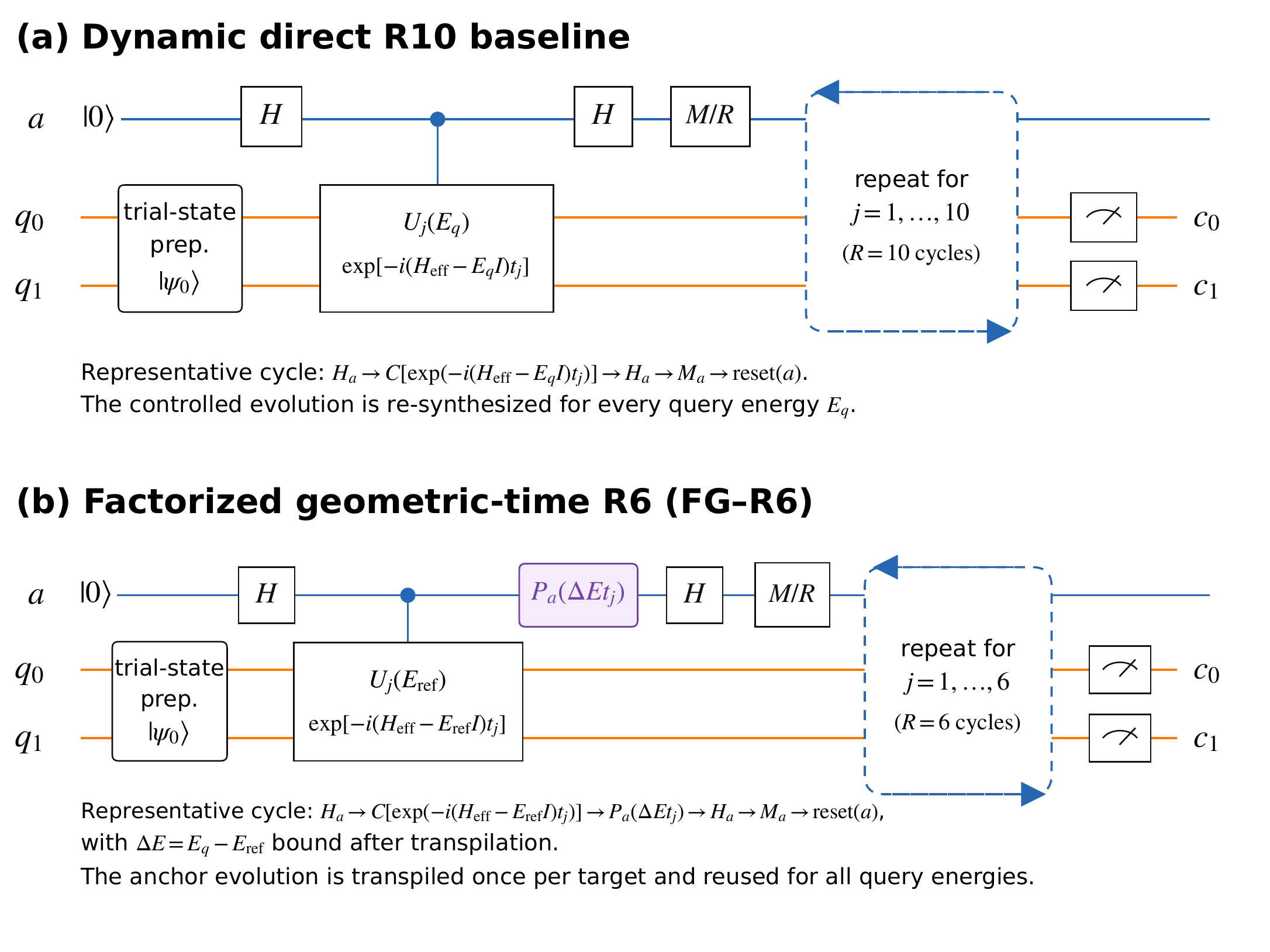}
  \caption{Representative cycles of the ancilla-recycled RODEO circuits.
  (a) Dynamic direct R10 uses ten cycles and compiles the
  query-dependent controlled evolution separately for each query energy.
  (b) \FG\ uses six geometric evolution times and a reference-energy
  circuit with the query-dependent ancilla phase bound after compilation.
  Both use two system qubits and one recycled ancilla.
  M/R denotes measurement after each cycle and reset before the next;
  no reset follows the final cycle.
  System qubits are measured only at the end, and classical ancilla
  records are omitted for clarity.
  The displayed cycle is included in the total of $R$ cycles;
trial-state preparation is performed only once.
Success requires all $R$ recorded ancilla outcomes to be zero,
irrespective of the terminal system bits $c_0,c_1$.
}
  \label{fig:circuit}
\end{figure*}

Here we retain the three-qubit register but shorten the evolution sequence using the six-cycle factorized geometric-time RODEO implementation (\FG), as shown in Fig.~\ref{fig:circuit}.
It combines the established separation of Hamiltonian evolution and
the query-energy phase \cite{Wang2024,Choi2021} with geometric
time sampling \cite{Patkowski2026}, using six evolution times
selected for the present scan window.
A reference-energy circuit is compiled once per target, with the query
offset subsequently bound to an ancilla phase.
Retrospective cycle-count and simulator tests support the six-cycle
choice within the adopted local spectral tolerances, without
establishing a minimum cycle count.

For each implementation, we estimate four positive-energy levels
at five trap strengths from raw IBM measurements, providing twenty
spectral inputs to the classical MERE analysis.
The four-dimensional Okubo--Lee--Suzuki (OLS) effective Hamiltonians are constructed classically from the reference eigensystems of the schematic $s$-wave $np$ interaction of Ref.~\cite{Wang2024}.
Exact diagonalization, the analytical scattering solution, and noiseless
and calibration-derived noisy simulations provide references.
Finite-shot energy uncertainties are propagated to the phase shift
separately from sensitivity to the extrapolation form.

Relative to dynamic R10, \FG\ reduces median compiled depth and
two-qubit-gate count by approximately one third.
For the present dataset and primary MERE fit, its maximum central
phase-shift deviation from exact-energy MERE is smaller on the
$0.1$--$30.0\MeV$ grid despite a larger root-mean-square (RMS)
trapped-energy deviation.
This gain arises mainly in the low-energy extrapolation region;
direct R10 has slightly smaller maximum and RMS phase residuals
on the $10.0$--$30.0\MeV$ common energy-interpolation subset.
The comparison links circuit compression to a nuclear continuum
observable and shows why spectral errors must be assessed through
the reconstruction and its energy range, rather than by their RMS alone.

\section{Trapped-spectrum formulation}
\label{sec:trapped_spectrum}

We consider a schematic $s$-wave $np$ square well of
Refs.~\cite{Wang2024,Mun2026}.
In the center-of-mass frame and in natural units ($\hbar=c=1$),
the relative Hamiltonian is
\begin{equation}
  \Htr(\omega)
  = T_{\mathrm{rel}}+V_{\mathrm{int}}(r)
  +\frac{1}{2}\mu\omega^2r^2,
  \label{eq:htrap}
\end{equation}
where $T_{\mathrm{rel}}=\boldsymbol{p}^{\,2}/(2\mu)$ is the relative kinetic energy and $\mu=469.460$ MeV is the neutron–proton reduced mass. 
The interaction is a square-well potential, $V_{\mathrm{int}}=-V_0$ for $r\leq W_0$ 
and zero otherwise, with $V_0=48.0002$ MeV and $W_0=1.70134$ fm.
The square well provides an analytical scattering reference.
Energies are measured from the free $np$ threshold, and the trap
strengths are $\omega=3.6,\,3.7,\,3.8,\,3.9,\,4.0\MeV$.

The reference spectrum is obtained by classical diagonalization in a
radial oscillator basis with $2n+\ell\leq N_{\max}=600$, 
corresponding to 301 basis states in the $s$-wave channel $\ell=0$.
The basis frequency $\Omega=60.0\MeV$ is distinct from the trap strength
$\omega$.
At each $\omega$, an OLS transformation
\cite{Okubo1954,SuzukiLee1980} yields a $4\times4$ effective
Hamiltonian preserving the four lowest positive eigenvalues of the
reference matrix to numerical precision.
Each reduced Hamiltonian is encoded on two system qubits. 
The twenty reference energies range from 8.855 to 35.340 MeV (see Table S1 in the Supplemental Material).

At each trapped eigenenergy $E_n(\omega)>0$, we evaluate
\cite{Luu2010,Zhang2020,Wang2024}
\begin{equation}
  \mathcal{K}_{\omega}(E_n)
  =-\sqrt{4\mu\omega}\,
  \frac{\Gamma\!\left(\frac{3}{4}-\frac{E_n}{2\omega}\right)}
       {\Gamma\!\left(\frac{1}{4}-\frac{E_n}{2\omega}\right)}.
  \label{eq:mere}
\end{equation}
With $p_n=\sqrt{2\mu E_n}$, the notation
$\mathcal{K}_{\omega}(E_n)=p_n\cot\delta_{0,\omega}(E_n)$
denotes a finite-confinement quantity, not the free-space effective-range
function, since the trap also perturbs the interaction region.
Numerical values of both $\mathcal{K}_{\omega}$ and the momenta in
$\mathrm{fm}^{-1}$ are obtained by dividing their natural-unit values
by $\hbar c=197.3269804\,\mathrm{MeV\,fm}$.

Following Ref.~\cite{Wang2024}, we fit the four points at each
$\omega$ by
\begin{equation}
  \mathcal{K}_{\omega}(E)
  \simeq d_0(\omega)+d_1(\omega)E+d_2(\omega)E^2.
  \label{eq:efit}
\end{equation}
Evaluating these five polynomials at a common scattering energy $E$,
we extrapolate their leading $\omega^2$ dependence to zero confinement:
\begin{equation}
  \begin{aligned}
    \mathcal{K}_{\omega}(E)
      &\simeq K_0(E)+K_2(E)\omega^2,\\
    \delta_0^{\mathrm{MERE}}(E)
      &=\frac{\pi}{2}
        -\arctan\!\left[\frac{K_0(E)}{p}\right].
  \end{aligned}
  \label{eq:omegafit}
\end{equation}
Here $p=\sqrt{2\mu E}>0$ in natural units and $K_0(E)$ estimates
$p\cot\delta_0(E)$, with $p$ and $K_0$ expressed in the same units.
The principal inverse tangent selects the continuous branch
$0<\delta_0<\pi$ used for both reconstructed and analytical phases.
Both stages use unweighted least squares and are repeated for the
energy-bootstrap samples; fit-form sensitivity is assessed separately.

Energy interpolation and trap removal are distinct.
Relative to the reference inputs, all five energy fits extrapolate
below $8.855\MeV$, while interpolation and extrapolation coexist
up to $9.957\MeV$.
Above this boundary, all five interpolate within the analysis range;
trap removal remains an extrapolation throughout.
On a uniform 300-point energy grid spanning $0.1$--$30.0\MeV$ in steps of $0.1$ MeV, the common interpolation subset therefore spans $10.0$--$30.0\MeV$ and contains 201 grid points.
Interval comparisons restrict the same reconstructed curves without
refitting.

Applying the primary prescription to the exact reference energies
defines $\delta_0^{\mathrm{ref}}$.
Its maximum absolute difference from the analytical square-well phase
$\delta_0^{\mathrm{an}}$ on the analysis grid is $0.1411^\circ$
below $8.855\MeV$ and $0.0502^\circ$ above it, decreasing to
$0.0372^\circ$ on the common-interpolation subset.
The full-grid maximum relative difference is $0.1128\%$.
These discrepancies characterize the fixed reference spectrum and
primary reconstruction, not quantum processing unit (QPU) uncertainties or a complete
extrapolation-error budget.

\section{Factorized geometric-time RODEO}
\label{sec:rodeo}

For $H=\Heff(\omega)$ and an initial state
$\ket{\psi_0}=\sum_\lambda c_\lambda\ket{\lambda}$, the ideal
probability of an all-zero ancilla record is \cite{Wang2024,Choi2021}
\begin{equation}
  P_R(\Eq)=\sum_\lambda |c_\lambda|^2
  f_{\mathbf t}(E_\lambda-\Eq),
  \label{eq:rodeo}
\end{equation}
where $f_{\mathbf t}(x)=\prod_{j=1}^{R}\cos^2(xt_j/2)$.
Both implementations use two system qubits and one recycled ancilla
\cite{Mun2026}.
The ancilla is measured after each cycle and reset only between cycles;
all-zero records define success.
The direct R10 reference uses ten fixed Gaussian-quantile times with
scale $\sigma=21.0\MeV^{-1}$ and separately compiles the controlled
$e^{-i(H-\Eq I)t_j}$ for each query energy.

The \FG\ circuit combines query-phase separation with a shorter
sequence (Fig.~\ref{fig:circuit}).
Using the established RODEO phase separation \cite{Wang2024,Choi2021},
\begin{equation}
  e^{-i(H-\Eq I)t_j}
  =e^{-i(H-\Eref I)t_j}e^{i\Delta E t_j},
  \label{eq:factorization}
\end{equation}
where $\Delta E=\Eq-\Eref$, the scalar factor becomes the ancilla phase
$P_a(\Delta E t_j)=\operatorname{diag}(1,e^{i\Delta E t_j})$.
The complete parameterized circuit is compiled once per target at fixed
$\Eref$, and $\Delta E$ is bound afterward for each query.
The identity is exact: at identical times, the numerical check gives
an absolute matrix-element difference no larger than
$3.56\times10^{-13}$.
This check is distinct from the approximation involved in changing
the time sequence.

Following geometric time sampling \cite{Patkowski2026}, we use
\begin{equation}
  t_j=t_{\max}\alpha^{-(j-1)},\qquad j=1,\ldots,6,
  \label{eq:times}
\end{equation}
with $t_{\max}=36.50539374\MeV^{-1}$ and $\alpha=1.199230706$.
At fixed $R=6$, these parameters were optimized against the isolated
R10 response on 1,001 points over $\pm50.0\keV$.
The maximum response difference is $8.51\times10^{-6}$, and
noiseless scans including all four levels give a maximum fitted-center
offset of $0.0853\keV$ in the $\pm40.0\keV$ query window.
The objective, bounds, and optimization settings and the full time sequences are provided in Sec. S3.2 and Table S4 of the Supplemental Material, respectively.
As will be discussed in Sec.~\ref{sec:results}, retrospective tests of the unchanged hardware sequence support this scan-specific choice but do not establish a minimum cycle count.

Each scan is centered on the classically known target energy $\Eref$.
We select the computational-basis state whose diagonal Hamiltonian
element is nearest $\Eref$, giving squared target overlaps of
0.855--0.922.
The seven offsets $0.0,\pm16.0,\pm28.0,\pm40.0\keV$, with the central query
repeated to check success-probability consistency, give
$20\times8\times2=320$ circuits at 500 shots each.
Both Direct R10 and \FG\ use 
$[q_{\rm sys0},q_{\rm sys1},q_a]=[114,119,113]$
on \texttt{ibm\_aachen}, without gate or measurement twirling,
runtime dynamical decoupling, or count debiasing.

The production controls comprise the fixed-time formula, ideal Aer
at 2,000 shots per circuit, and calibration-derived noisy Aer at
500 shots \cite{JavadiAbhari2024}; the latter only approximates
processor noise.
The additional cycle-count study, detailed in Secs. S8 and S9 of the Supplemental Material, uses separate noiseless and synthetic-noise simulations. 
No additional IBM measurements were performed for the retrospective
cycle-count study.

Each isolated peak is fitted by binomial maximum likelihood to
\begin{equation}
  P_{\rm fit}(\Eq)=K f_{\mathbf t}(\Ec-\Eq),
  \label{eq:peakfit}
\end{equation}
where $\Ec$ estimates the trapped energy and $K$ is a fitted
probability scale, not a state fidelity.
Repeated central counts are pooled without drift correction.
We generate 2,000 parametric binomial replicas per energy fit,
refitted by approximate profile weighted least squares, and independently sample the twenty centers in 1,500 MERE replicas.
These finite-shot uncertainties are conditional on the peak model and exclude cross-target covariance and run-to-run drift; details of the fitting and uncertainty propagation are given in Secs. S5 and S6 of the Supplemental Material.
\begin{figure}[t]
  \centering
  \includegraphics[width=1.00\columnwidth]{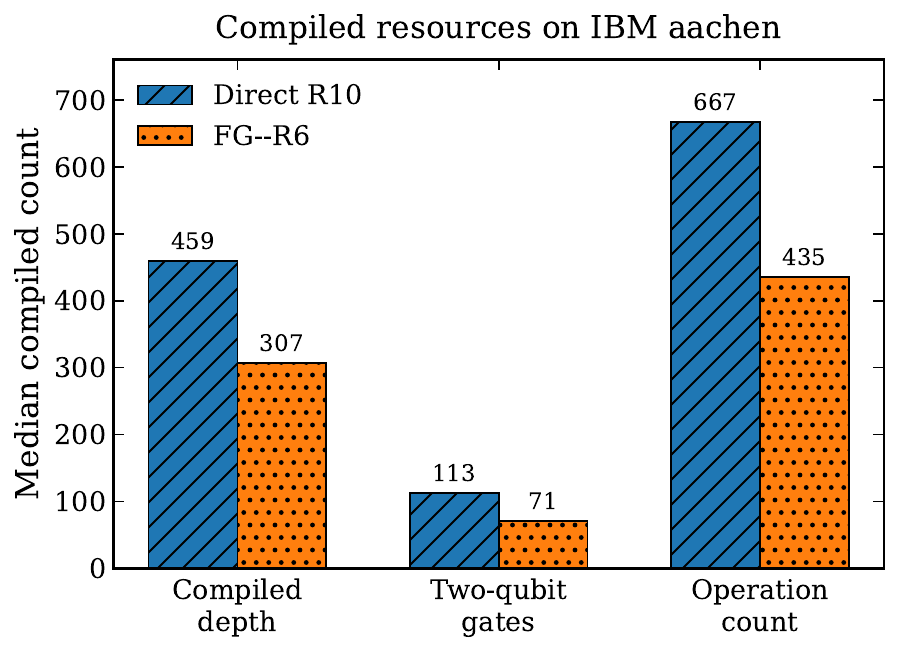}
\caption{
Compiled resources on \texttt{ibm\_aachen} for dynamic direct R10 and FG--R6.
Circuit depth, two-qubit-gate count, and total operation count are shown as
medians over 160 circuits per implementation after compilation to the backend
instruction set architecture (ISA).
Both implementations use three active qubits.
}
  \label{fig:resources}
\end{figure}

\section{Results}
\label{sec:results}

All 320 circuit results were retrieved without missing or unexpected
identifiers, yielding twenty fitted levels and 80,000 shots per method.
As shown in Fig.~\ref{fig:resources}, 
$\FG$ reduces the median compiled depth from 459 to 307 and the median two-qubit-gate count from 113 to 71 relative to direct R10, corresponding to reductions of $33.1\%$ and $37.2\%$, respectively.
Both implementations use three active qubits \cite{Mun2026}; the
reduction is from ten controlled evolutions to six, not in circuit width.
Mid-circuit measurement and reset still introduce latency and errors
\cite{DeCross2023}.

Retrospective cycle-count tests support the six-cycle choice.
With the original optimization objective, the tested $R=4$ and $R=5$
candidate sequences exceed the adopted $0.10\keV$ noiseless center-error
tolerance, whereas $R=6$ satisfies it, including the tested scan-center
shifts of up to $10.0\keV$ (maximum error $0.0926\keV$).
Its isolated response matches R10 within $8.51\times10^{-6}$ on the
matching grid.
An $R=4$ sequence separately retuned using the known spectrum also
passes the local tests; \FG\ is therefore a validated accuracy--resource
compromise, not a minimum-cycle solution.
The cycle-count, off-target, and simulator comparisons are given in Secs. S8 and S9 of the Supplemental Material.

Table~\ref{tab:ibm_summary} separates spectral and scattering metrics.
At equal hardware shot budgets, \FG\ gives a larger median fitted
amplitude and smaller finite-shot energy uncertainty, but its RMS
center offset $\Delta E_c=\Ec-\Eref$ increases from $0.793$ to
$1.648\keV$.
Improved precision therefore does not imply improved absolute energy
accuracy.
Repeated central queries differ by more than two nominal binomial
standard errors for 11 of 20 direct-R10 targets and 1 of 20
\FG\ targets (Table S3 of the Supplemental Material).
No drift correction or uncertainty rescaling is applied.
The original ideal-Aer controls give RMS offsets of $0.219$ and
$0.132\keV$ at 2,000 shots per circuit, while calibration-derived
noisy Aer gives $0.491$ and $0.493\keV$ at 500 shots.
The latter does not reproduce the larger hardware offsets.
These data-source comparisons, the finite-confinement MERE inputs,
and representative energy and trap-removal fits are provided in 
Table S6, Fig. S2, and Figs. S3 and S4 of the Supplemental Material, respectively.

\begin{table}
  \centering
  \caption{Raw IBM benchmark: twenty trapped levels and 80,000
  shots per method.
  Here $\Delta\delta_0=\delta_0^{\mathrm{QPU}}-\delta_0^{\mathrm{ref}}$,
  where $\delta_0^{\mathrm{ref}}$ is exact-energy MERE with the primary
  fit, and $\delta_0^{\mathrm{an}}$ is the analytical square-well phase.
  Phase entries are in degrees.
 For phase quantities, maxima and RMS values are evaluated on an energy grid with 0.1 MeV spacing: 300 points over the full interval (0.1--30.0 MeV) and 201 over the common interpolation subset (10.0--30.0 MeV). Trap removal remains an extrapolation.
  Standard deviations describe finite-shot uncertainty only.
  $h_{\mathrm{model}}$ is the maximum pointwise half-range over six
  fit forms, not a confidence interval.}
  \label{tab:ibm_summary}

  \small
  \setlength{\tabcolsep}{4pt}
  \renewcommand{\arraystretch}{1.04}
  \vspace{5pt}
  \begin{tabular}{@{}lrr@{}}
    \toprule
    Quantity & Direct R10 & FG--R6 \\
    \midrule

    \multicolumn{3}{@{}l}{
      \textit{Trapped-energy estimates: twenty levels}} \\
    RMS $\Delta E_c$ (keV)
      & 0.793 & 1.648 \\
    Median $\sigma_E$ (keV)
      & 0.610 & 0.534 \\
    Median $K$
      & 0.537 & 0.652 \\

    \midrule
    \multicolumn{3}{@{}l}{
      \textit{Phases: $E=0.1$--$30.0$} MeV} \\
    $\max_E|\Delta\delta_0|$
      & 0.842 & 0.163 \\
    $\mathrm{RMS}_E(\Delta\delta_0)$
      & 0.305 & 0.137 \\
    $\max_E|\delta_0^{\mathrm{QPU}}-\delta_0^{\mathrm{an}}|$
      & 0.981 & 0.156 \\
    $\max_E\sigma_\delta$
      & 0.429 & 0.401 \\
    $h_{\mathrm{model}}$
      & 20.12 & 6.75 \\

    \midrule
    \multicolumn{3}{@{}l}{
      \textit{Phases: $E=10.0$--$30.0$} MeV} \\
    $\max_E|\Delta\delta_0|$
      & 0.137 & 0.159 \\
    $\mathrm{RMS}_E(\Delta\delta_0)$
      & 0.102 & 0.129 \\
    $\max_E|\delta_0^{\mathrm{QPU}}-\delta_0^{\mathrm{an}}|$
      & 0.160 & 0.156 \\
    $\max_E\sigma_\delta$
      & 0.074 & 0.065 \\
    $h_{\mathrm{model}}$
      & 3.19 & 1.12 \\

    \bottomrule
  \end{tabular}
\end{table}

\begin{figure*}
  \centering
  \includegraphics[width=\textwidth]{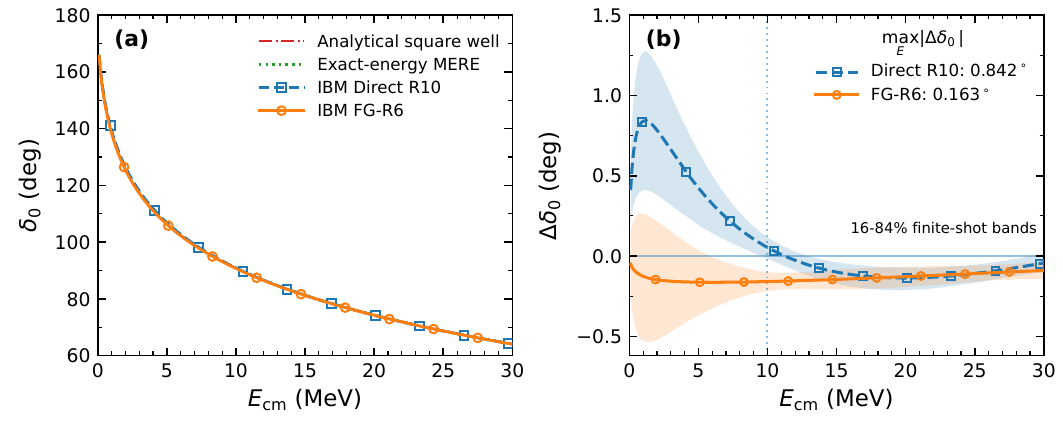}
  \caption{Free-space phase shifts reconstructed from raw IBM measurements.
  (a) Direct R10 and \FG\ results from twenty trapped levels per method,
  compared with the analytical square-well solution and exact-energy MERE.
  (b) Residuals relative to the exact-energy MERE reference,
  $\Delta\delta_0=\delta_0^{\mathrm{QPU}}-\delta_0^{\mathrm{ref}}$.
  Shading denotes pointwise 16--84\% finite-shot bootstrap intervals,
  excluding fit-form sensitivity, cross-target covariance, and run-to-run
  drift.
  Symbols identify reconstructed curves, not independent continuum
  measurements.
  The dotted line at $E=9.9566\MeV$ marks the onset of common energy
  interpolation; trap removal remains an extrapolation throughout.
  Maxima use $E=0.1,0.2,\ldots,30.0\MeV$.
 }
  \label{fig:phase}
\end{figure*}

Figure~\ref{fig:phase}(a) compares the two IBM reconstructions with the analytical solution and the exact-energy MERE reference. Figure~\ref{fig:phase}(b) shows the corresponding residuals $\Delta\delta_0$.
On the full $0.1$--$30.0\MeV$ grid, the maximum absolute residual
decreases from $0.842^\circ$ to $0.163^\circ$, an $80.6\%$ reduction,
and the phase RMS residual decreases from $0.305^\circ$ to
$0.137^\circ$.
The two maximum residuals occur at $1.2$ and $6.1\MeV$, respectively,
below the lowest trapped input.
The maximum deviation from the analytical solution also decreases,
from $0.981^\circ$ to $0.156^\circ$.

The full-grid gain is driven by the larger low-energy deviation of
R10, not by uniformly improved phase shifts.
On the $10.0$--$30.0\MeV$ common-interpolation subset, direct R10
has smaller maximum and RMS residuals relative to exact-energy MERE:
$0.137^\circ$ and $0.102^\circ$, compared with $0.159^\circ$ and
$0.129^\circ$ for \FG.
The maximum deviations from the analytical solution are instead
similar, $0.160^\circ$ and $0.156^\circ$.
These subset metrics restrict the same reconstructed curves;
no refitting or additional continuum measurements are involved.

The full-grid reversal shows that the RMS spectral error alone need
not rank the accuracy of a reconstructed scattering observable.
All twenty level shifts enter both the energy and trap-strength fits
in Eqs.~\eqref{eq:efit} and \eqref{eq:omegafit}.
The level-resolved offsets listed in Table S7 of the Supplemental Material are predominantly positive for \FG\ 
and sign-changing for direct R10.
This pattern does not establish an error-cancellation mechanism or
identify the hardware noise responsible for the simulator--QPU difference.

Extrapolation sensitivity remains appreciable.
Across the six combinations of energy-polynomial orders one to three
and trap fits with or without an $\omega^4$ term, the maximum
half-range decreases from $20.12^\circ$ to $6.75^\circ$ on the full
grid and from $3.19^\circ$ to $1.12^\circ$ on the common-interpolation
subset.
This spread is a fit-form diagnostic, not a calibrated truncation
uncertainty or a finite-shot interval.
All energy fits extrapolate below $8.855\MeV$, and trap removal
remains an extrapolation at every energy.
The finite-shot bands omit cross-target covariance and run-to-run drift.
Smaller central residuals therefore do not establish a corresponding
reduction in total uncertainty.

The calculation extends the simulator study of Ref.~\cite{Wang2024}
and our single-level static--dynamic benchmark \cite{Mun2026} to
complete hardware-derived spectral inputs for classical free-space
phase-shift reconstruction, complementing quantum-subspace approaches
\cite{Sharma2024}.
The resource reduction and scattering results validate the combined
\FG\ implementation in the present four-dimensional OLS benchmark,
not quantum advantage or many-body scalability.
The hardware data do not separately determine the effects of phase
compilation and time sampling, or establish independent-run reproducibility.

\section{Conclusions}
\label{sec:conclusions}

We have reconstructed elastic $s$-wave phase shifts for a schematic
$np$ interaction by combining IBM spectral measurements with classical
MERE extrapolation.
Each implementation supplies four positive-energy levels at five trap
strengths from classically constructed four-dimensional OLS Hamiltonians,
extending our single-level ancilla-recycling benchmark \cite{Mun2026}
to all twenty spectral inputs for free-space scattering.

The \FG\ circuit retains three active qubits and combines exact
query-phase separation, post-compilation binding, and six geometric
evolution times.
Retrospective numerical tests support this choice within the adopted
local spectral tolerances, without establishing a minimum cycle count
or universal optimum.
Relative to dynamic direct R10, it reduces median compiled depth
and two-qubit-gate count by $33.1\%$ and $37.2\%$, respectively,
with a larger median fitted amplitude and smaller median finite-shot
energy uncertainty.

For the present dataset and primary fit on the $0.1$--$30.0\MeV$ grid,
the maximum central phase-shift deviation from exact-energy MERE
decreases from $0.842^\circ$ to $0.163^\circ$, despite a larger
RMS energy offset underestimated by the calibration-derived simulator.
The full-grid gain arises mainly in the low-energy extrapolation region.
On the $10.0$--$30.0\MeV$ common energy-interpolation subset, direct R10
instead has slightly smaller maximum and RMS residuals relative to
the same reference.
The RMS spectral error alone therefore need not rank scattering
accuracy, which also depends on the reconstruction and energy interval.

A natural next step is to extend the present $s$-wave benchmark to 
resonant higher-partial-wave scattering. Neutron--$\alpha$ scattering 
provides a particularly suitable case because its low-energy 
$P_{3/2}$ and $P_{1/2}$ channels generate the unbound 
$J^\pi=3/2^-$ ground-state and $1/2^-$ excited-state resonances of 
${}^{5}\mathrm{He}$, whose phase shifts and resonance parameters are 
well constrained by experimental $R$-matrix analyses 
\cite{Nollett2007,Tilley2002}. Reconstructing these two channels would 
test whether the trapped-spectrum method can recover not only phase 
shifts but also resonance energies, widths, and the associated 
spin--orbit splitting. This extension requires channel-resolved 
trapped Hamiltonians and the higher-partial-wave effective-range 
function 
$p^{2\ell+1}\cot\delta_\ell$,  
while retaining the same RODEO-based 
spectral-estimation framework. A subsequent application to 
proton--$\alpha$ scattering would introduce the long-range Coulomb 
interaction and require a Coulomb-modified trap quantization 
condition. Comparing the $n\alpha$ and $p\alpha$ systems would then 
provide a controlled route from the present short-range $s$-wave 
demonstration to spin-dependent, resonant, and Coulomb-distorted 
nuclear scattering on quantum hardware.

Fit-form sensitivity remains appreciable, and finite-shot bands
exclude cross-target covariance and run-to-run drift.
These central-value comparisons do not establish a reduction in
total uncertainty.
Separate hardware controls and independent repetitions would clarify
the contributions of compilation, time sampling, and execution variability.
The present benchmark thus provides a starting point for these
extensions to larger model spaces, realistic interactions, and
few-body reaction dynamics.

\section*{Data Availability}
The raw QPU counts, job identifiers, transpiled circuits,
processed data, and analysis scripts supporting this work
are available from the corresponding authors upon reasonable request.

\section*{Acknowledgments}
This work was supported by the National Research Foundation of Korea (NRF) through grants funded by the Korean government (Ministry of Science and ICT, MSIT): Grant Nos. RS-2026-25487837 and RS-2018-NR031074 (M.-H.M.); Grant Nos. RS-2021-NR060129, RS-2024-00460031, and RS-2025-16071941 (M.-K.C.); and Grant No. RS-2025-00513410 (E.-J.H.). Additional support was provided through NRF Grant Nos. RS-2025-24533596 and RS-2025-25400847, funded by MSIT and the Ministry of Education, respectively (J.-B.P.).
Quantum computing cloud resources were provided by the Korea Institute of Science and Technology Information (KISTI) through an NRF grant funded by the Korean government (MSIT) (Grant No. RS-2025-24534879).
This research was also supported by the “Quantum Information Science R\&D Ecosystem Creation” program through the NRF, funded by the Korean government (MSIT) (Grant No. 2020M3H3A1110365).


\end{document}


\begin{frontmatter}
\title{Supplemental Material for ``Nuclear scattering phase shifts with
factorized geometric-time RODEO on a quantum processor''}
\author[knu]{Myeong-Hwan Mun}
\author[ssu]{Jubin Park\corref{cor1}}
\ead{honolov@ssu.ac.kr}
\author[ssu]{Myung-Ki Cheoun\corref{cor2}}
\ead{cheoun@ssu.ac.kr}
\author[hu]{Eunja Ha}
\cortext[cor1]{Corresponding author}
\cortext[cor2]{Corresponding author}
\address[knu]{Department of Physics, Kyungpook National University,
Daegu 41566, Republic of Korea}
\address[ssu]{Department of Physics and Origin of Matter and Evolution
of Galaxies Institute, Soongsil University, Seoul 06978, Republic of Korea}
\address[hu]{Department of Physics and Research Institute for Natural
Sciences, Hanyang University, Seoul 04763, Republic of Korea}
\end{frontmatter}

The production dataset comprises the original IBM measurements
for the dynamic direct ten-cycle RODEO implementation (direct R10)
and the six-cycle factorized geometric-time implementation (FG--R6),
together with their local and simulator controls.
The retrospective 
cycle study adds classical and simulator tests only: it neither retunes
the archived hardware R6 times nor replaces measured energies.

\section{Hamiltonian and reference trapped spectrum}
\label{sec:hamiltonian}
The schematic $s$-wave neutron--proton ($np$) square well of Refs.~\cite{Wang2024,Mun2026}
provides an analytical benchmark, not a realistic interaction.
With $\hbar=c=1$ and $\mu=469.460\MeV$,
\begin{align}
 \Htr(\omega)&=\frac{\boldsymbol p^{\,2}}{2\mu}
       +V_{\mathrm{int}}(r)+\tfrac12\mu\omega^2r^2,\label{eq:Strap}\\
 V_{\mathrm{int}}(r)&=
 \begin{cases}-V_0,&r\leq W_0,\\0,&r>W_0,\end{cases}\label{eq:Swell}\\
 V_0&=48.0002\MeV,\qquad W_0=1.70134\fm.\nonumber
\end{align}
The trap acts only on relative motion; energies are measured from the
free $np$ threshold. Classical diagonalization uses $2n+\ell\leq600$
(301 $s$-wave states), basis frequency $\Omega=60.0\MeV$, and 500-point
Gauss--Legendre quadrature over the well. The external frequencies
$\omega=3.6,3.7,3.8,3.9,4.0\MeV$ are distinct from $\Omega$.

Let $A$ contain projections of the four lowest positive-energy
reference eigenvectors onto the first four oscillator states. The
hermitian Okubo--Lee--Suzuki (OLS) construction \cite{Okubo1954,SuzukiLee1980} is
\begin{equation}
 B=A(A^\dagger A)^{-1/2},\qquad
 \Heff=B\,\operatorname{diag}(E_1,E_2,E_3,E_4)B^\dagger.
 \label{eq:Sols}
\end{equation}
It preserves those finite-reference-space energies (Table~\ref{tab:levels})
and is encoded in two system qubits. The original $\omega=3.7\MeV$
matrix agrees with its independent archive within
$8.68\times10^{-10}\MeV$ per element; this reconstruction check is not
an infinite-basis uncertainty or a quantum solution of the 301-state problem.

\begin{table}[!htbp]
  \centering
  \caption{Positive reference energies (MeV), retained by the four-dimensional OLS Hamiltonians. Displayed values are rounded.}
  \label{tab:levels}
  \footnotesize
  \setlength{\tabcolsep}{2.1pt}
  \begin{tabular}{@{}crrrr@{}}
    \toprule
    $\omega$ (MeV) & $E_1$ & $E_2$ & $E_3$ & $E_4$ \\
    \midrule
    3.6 & 8.855470 & 16.643000 & 24.213795 & 31.691304 \\
    3.7 & 9.129551 & 17.134039 & 24.916082 & 32.602105 \\
    3.8 & 9.404443 & 17.625943 & 25.619273 & 33.513827 \\
    3.9 & 9.680129 & 18.118693 & 26.323347 & 34.426445 \\
    4.0 & 9.956592 & 18.612272 & 27.028285 & 35.339939 \\
    \bottomrule
  \end{tabular}
\end{table}

With lengths in fm, define
\begin{equation}
 p(E)=\frac{\sqrt{2\mu E}}{\hc},\qquad
 k(E)=\frac{\sqrt{2\mu(E+V_0)}}{\hc},
 \label{eq:Smomenta}
\end{equation}
where $\hc=197.3269804\,\mathrm{MeV\,fm}$. Matching the radial waves gives
$k\cot(kW_0)=p\cot[pW_0+\delta_0^{\rm an}]$ and
\begin{equation}
 \delta_0^{\rm an}(E)={} \arg[k\cos(kW_0)+ip\sin(kW_0)]
 -pW_0+Q(E)\,\pi.~
 \label{eq:Sanalytic}
\end{equation}
The quadrant-aware argument equals the recorded two-argument
arctangent. $Q$ fixes the shared continuous branch $0<\delta_0<\pi$,
not a correction to quantum processing unit (QPU) data. Reported phases are in degrees.

\section{Trial states and fixed query grid}
\label{sec:trials}
The implemented \emph{diagonal-nearest} rule is
\begin{equation}
 b_n=\underset{b=0,1,2,3}{\arg\min}|(\Heff)_{bb}-E_n|,
 \qquad w_n=|\langle n|b_n\rangle|^2.
 \label{eq:Strial}
\end{equation}
In the ordered basis $\{\ket{00},\ket{01},\ket{10},\ket{11}\}$,
the $\ket{q_1q_0}$ choices are $00,11,10,01$ for levels 1--4 at
every trap. These are OLS labels, not nucleon occupations; the rule
does not explicitly maximize exact overlap (Table~\ref{tab:overlaps}).

\begin{table}[!htbp]
  \centering
  \caption{Squared target overlaps for the diagonal-nearest trial states $\ket{q_1q_0}$; these bits label OLS basis states, not nucleon occupations.}
  \label{tab:overlaps}
  \footnotesize
  \setlength{\tabcolsep}{2.1pt}
  \begin{tabular}{@{}crrrr@{}}
    \toprule
    $\omega$ (MeV) & $E_1$: $\ket{00}$ & $E_2$: $\ket{11}$ & $E_3$: $\ket{10}$ & $E_4$: $\ket{01}$ \\
    \midrule
    3.6 & 0.897683 & 0.883041 & 0.855202 & 0.867051 \\
    3.7 & 0.902214 & 0.894193 & 0.867914 & 0.871462 \\
    3.8 & 0.906169 & 0.904422 & 0.880093 & 0.875233 \\
    3.9 & 0.909560 & 0.913712 & 0.891722 & 0.878369 \\
    4.0 & 0.912403 & 0.922046 & 0.902786 & 0.880876 \\
    \bottomrule
  \end{tabular}
\end{table}

The requested sequence around the known $\Eref=E_n(\omega)$ is
\begin{equation}
 \Delta E_q/\keV=0,+16,-16,+28,-28,+40,-40,0.
 \label{eq:Squeries}
\end{equation}
The central query is intentionally repeated to test 
success-probability consistency.
At 500 shots per occurrence, the eight entries give 4,000 shots per
target and 80,000 per method. Pooling the repeated center gives seven
fit points: 1,000 center shots and 500 elsewhere. This is a local scan,
not blind spectral discovery. For the two center occurrences,
\begin{equation}
 \begin{aligned}
 \Delta\widehat P_0&=\widehat P_{0,2}-\widehat P_{0,1},\\
 s_{\Delta P}^2&=\sum_{a=1}^{2}
   \widehat P_{0,a}(1-\widehat P_{0,a})/N_a,\qquad
 z=\Delta\widehat P_0/s_{\Delta P}.
 \end{aligned}\label{eq:Srepeat}
\end{equation}
Table~\ref{tab:centerrepeat}, checked against counts, shows discrepancies
beyond binomial sampling for several pairs. No drift correction or error
rescaling is applied: the shot band remains conditional on pooled
probabilities. These are probability-consistency checks, not separate
energy fits or timing-resolved drift measurements. Requested order does
not establish execution chronology, and the batches in
Ref.~\cite{Mun2026} are not repeats of this twenty-level campaign.

\begin{table}[!htbp]
 \centering
 \caption{Original IBM repeated-center consistency over twenty targets per method. The $|z|$ counts are descriptive, with no multiple-comparison correction.}
 \label{tab:centerrepeat}
 \small
 \setlength{\tabcolsep}{4pt}
 \begin{tabular}{@{}lrr@{}}
  \toprule
  Quantity & Direct R10 & FG--R6 \\
  \midrule
  Median $|\Delta\widehat P_0|$ & 0.064 & 0.036 \\
  Maximum $|\Delta\widehat P_0|$ & 0.106 & 0.102 \\
  Maximum $|z|$ & 3.373 & 3.314 \\
  Targets with $|z|>2$ & 11/20 & 1/20 \\
  Targets with $|z|>3$ & 2/20 & 1/20 \\
  \bottomrule
 \end{tabular}
\end{table}

\section{RODEO schedules, factorization, and numerical design}
\label{sec:schedules}
For $\ket{\psi_0}=\sum_\lambda c_\lambda\ket{\lambda}$, the ideal response is
\cite{Choi2021,Wang2024}
\begin{equation}
 P_{\mathbf t}(\Eq)=\sum_\lambda|c_\lambda|^2 f_{\mathbf t}(E_\lambda-\Eq),
 \quad f_{\mathbf t}(x)=\prod_{j=1}^{R}\cos^2(xt_j/2).
 \label{eq:Sfilter}
\end{equation}
The R10 times are fixed standard-normal quantiles with the scale of
Ref.~\cite{Wang2024}, not fresh Gaussian draws:
\begin{equation}
 t_j^{(10)}=21\,\Phi^{-1}[(j-1/2)/10]\,\mathrm{MeV}^{-1},
 \quad j=1,\ldots,10.
 \label{eq:SR10}
\end{equation}
The unchanged production R6 configuration is
\begin{equation}
 \begin{aligned}
 t_j^{(6)}&=t_{\max}\alpha^{-(j-1)},\quad j=1,\ldots,6,\\
 t_{\max}&=36.5053937358\,\mathrm{MeV}^{-1},\quad
 \alpha=1.1992307057.
 \end{aligned}\label{eq:SR6}
\end{equation}
Table~\ref{tab:times} lists both sequences, used for all targets.

\begin{table}[!htbp]
  \centering
  \caption{Fixed production times ($\mathrm{MeV}^{-1}$), rounded for display. Equations~\eqref{eq:SR10} and \eqref{eq:SR6} define the sequences; no fresh random-time averaging is used.}
  \label{tab:times}
  \small
  \setlength{\tabcolsep}{4pt}
  \begin{tabular}{@{}crr@{}}
    \toprule
    Cycle $j$ & Direct R10 & FG--R6 \\
    \midrule
    1 & -34.54192616598 & 36.50539373580 \\
    2 & -21.76510117937 & 30.44067631215 \\
    3 & -14.16428475412 & 25.38350308032 \\
    4 & -8.09172979456 & 21.16648861613 \\
    5 & -2.63888828396 & 17.65005558608 \\
    6 & 2.63888828396 & 14.71781493101 \\
    7 & 8.09172979456 & -- \\
    8 & 14.16428475412 & -- \\
    9 & 21.76510117937 & -- \\
    10 & 34.54192616598 & -- \\
    \bottomrule
  \end{tabular}
\end{table}

\subsection{Exact phase separation and circuit records}
Controlled evolutions use numerical exponentials of $\Heff$, with no
product-formula approximation to the $4\times4$ matrix. The established
phase separation \cite{Wang2024,Choi2021} is
\begin{equation}
 e^{-i(H-\Eq I)t_j}=e^{-i(H-\Eref I)t_j}e^{i(\Eq-\Eref)t_j}.
 \label{eq:Sfactor}
\end{equation}
The scalar becomes a relative ancilla phase
$P_a(\Delta Et_j)=\operatorname{diag}(1,e^{i\Delta Et_j})$.
FG--R6 compiles the complete target circuit with $\Delta E=\Eq-\Eref$
unbound, then assigns it per query; direct R10 recompiles per query.
At identical times the check is
\begin{equation}
 \epsilon_{\rm fact}=
 \max_{\mathrm{tasks},j,m,n}|(U_j^{\rm dir}-U_j^{\rm fact})_{mn}|
 =3.56\times10^{-13}.
 \label{eq:Snorm}
\end{equation}
This is an elementwise discrepancy, not a spectral norm, hardware
fidelity, or equivalence of R10 and R6.

Both circuits use two system qubits and one ancilla. Each cycle applies
Hadamard--controlled evolution--query phase (where separate)--Hadamard--
ancilla measurement, storing a fresh classical bit. There are $R$
measurements and $R-1$ resets, with terminal system readout after the last
cycle. All entries execute fully; success is an all-zero ancilla record,
irrespective of the terminal system bits.

\subsection{Optimization at fixed cycle count}
\label{sec:objective}
For geometric sampling \cite{Patkowski2026}, set $\theta=(t_{\max},\alpha)$,
$x_k=(-0.050+0.0001k)\MeV$ ($k=0,\ldots,1000$), and
$D_k=f_{\mathbf t(\theta)}(x_k)-f_{\mathbf t^{(10)}}(x_k)$. The loss is
\begin{equation}
 \begin{split}
 \mathcal L_R(\theta)={}&1001^{-1}\sum_kD_k^2+10\max_k|D_k|^2\\
 &+10^{-7}\!\left[\frac{T_R-0.9T_{10}}{1\,\mathrm{MeV}^{-1}}\right]^2,
 \quad T_R=\sum_j|t_j|.
 \end{split}\label{eq:Sloss}
\end{equation}
Bounds are $10\leq t_{\max}/(\mathrm{MeV}^{-1})\leq70$ and
$1.01\leq\alpha\leq3$. Differential evolution uses seed 20,260,812,
population multiplier 15, tolerance $10^{-11}$, 1200 maximum iterations,
and final polishing. The time-sum penalty acts on both sides of its
chosen value; it models neither QPU duration nor a physical resolution
limit. MeV detunings ensure dimensionless $xt_j$.

Production optimizes and stores parameters at \emph{fixed} $R=6$,
without searching $R$. The loss uses no off-target eigenenergies;
the full spectrum enters subsequent validation with inherited limits
\begin{equation}
 D_{\max}\leq2\times10^{-3},\qquad
 \max_{\omega,n}|E^{\rm ideal}_{c,n}-E_n|\leq0.10\keV.
 \label{eq:Stolerances}
\end{equation}
Stored R6 gives $D_{\max}=8.51\times10^{-6}$ on the matching grid and
maximum four-level center error $0.0853\keV$ over the actual $\pm40.0\keV$
queries. These are local design tests, not universal nuclear tolerances;
Sec.~\ref{sec:cycle} gives the retrospective cycle comparison.

\section{Original IBM execution and simulation controls}
\label{sec:execution}
\begin{table}[!htbp]
  \centering
  \caption{Original production metadata. Resource medians use 160 circuits per method; software versions describe the recorded environment.}
  \label{tab:execution}
  \footnotesize
  \setlength{\tabcolsep}{4pt}
  \begin{tabular}{@{}ll@{}}
    \toprule
    Item & Recorded value \\
    \midrule
    Backend & \texttt{ibm\_aachen} \\
    Instance & \texttt{0\_2026\_HU-eu} \\
    Layout $(q_0,q_1,a)$ & $[114,119,113]$ \\
    Targets / query entries / methods & $20/8/2$ \\
    QPU circuits / shots per circuit & $320/500$ \\
    Shots per method / total shots & $80\,000/160\,000$ \\
    Retrieved task identifiers & $320/320$ \\
    Median depth: R10 / FG--R6 & $459/307$ \\
    Median two-qubit gates: R10 / FG--R6 & $113/71$ \\
    Median size: R10 / FG--R6 & $667/435$ \\
    Qiskit / Aer / IBM Runtime & $2.5.1/0.17.2/0.48.0$ \\
    Compilation level / seed & $3/20260816$ \\
    \bottomrule
  \end{tabular}
\end{table}

Both production methods share Table~\ref{tab:execution}'s layout and
shot budget. Gate/measurement twirling, Runtime dynamical decoupling,
fractional-gate mode, and count debiasing are off. 
The submitted circuits are archived in Qiskit's portable binary 
QPY format and retain the backend-specific mid-circuit 
measurement instruction.
Only temporary Aer copies
replace \texttt{measure\_2} with standard measurement while retaining
calibration-derived noise; this does not reproduce the full measurement
channel or timing effects.

\emph{Local formula} gives exact reduced-Hamiltonian probabilities;
\emph{local sampled} draws 2000 binomial shots per circuit; \emph{ideal
Aer} is noiseless gate-level simulation at 2000 shots; \emph{calibration
Aer} uses the archived IBM noise model at 500 shots; and \emph{IBM QPU}
means 500 raw Sampler shots \cite{JavadiAbhari2024}. 
The two IBM implementations have equal shot budgets;
the shot counts are not uniform across data sources.
Agreement or disagreement with approximate Aer noise does not identify
coherent, crosstalk, drift, or reset mechanisms.

\section{Peak fitting, repeated queries, and bootstrap}
\label{sec:peak}
The full response (\ref{eq:Sfilter}) is approximated near one target by
\begin{equation}
 P_{\rm fit}(\Eq)=K f_{\mathbf t}(\Ec-\Eq),
 \label{eq:Speak}
\end{equation}
without an additive background. $K$ is a probability scale, not state
fidelity. 
At query energy $x_i=E_{q,i}$, let $k_i$ denote the pooled
all-zero ancilla count from $N_i$ shots.
The binomial maximum-likelihood estimate (MLE) minimizes
\begin{equation}
 -\log\mathcal L=-\sum_i[k_i\log p_i+(N_i-k_i)\log(1-p_i)],
 \label{eq:Slike}
\end{equation}
with $p_i=Kf_{\mathbf t}(x_i-\Ec)$, omitting constant binomial factors.
Bounds are $\Ec\in\Eref\pm25.0\keV$ and $K\in[10^{-6},0.999999]$.
L-BFGS-B uses 5000 maximum iterations, tolerance $10^{-14}$, and
probability clipping to $[10^{-10},1-10^{-10}]$ in logarithms.
$\Delta E_c=\Ec-\Eref$; the Pearson diagnostic has five nominal degrees
of freedom (seven points, two parameters), with no uncertainty rescaling.

\begin{figure*}[!tbp]
 \centering
 \suppgraphic{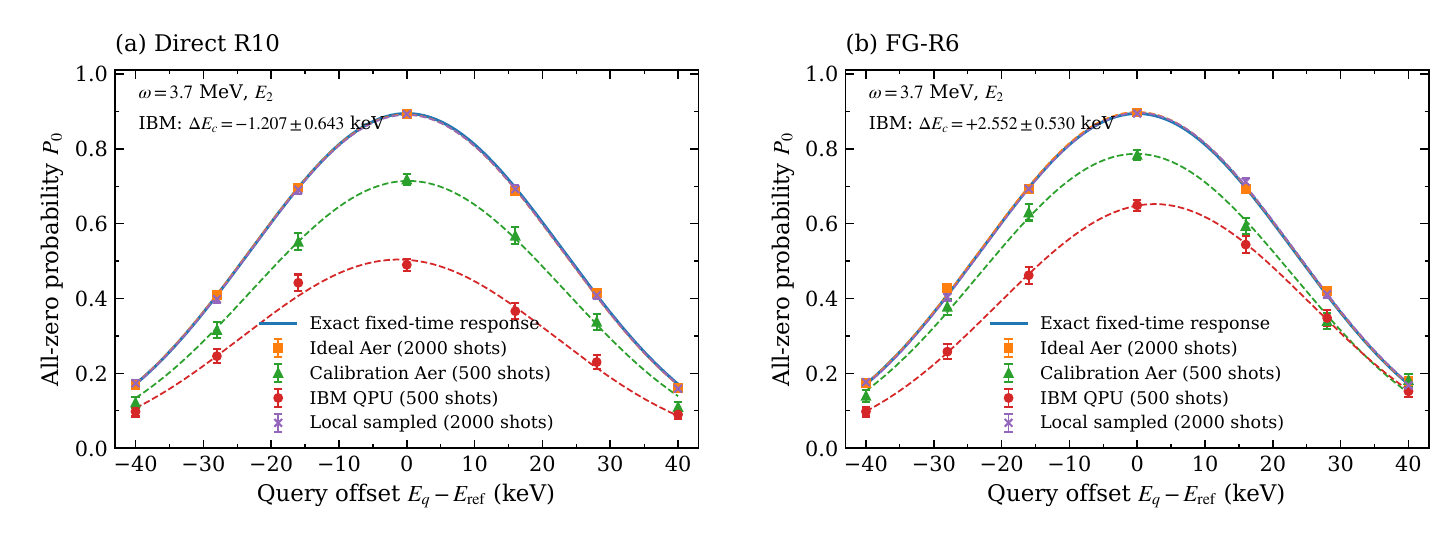}{0.98\textwidth}{49mm}
 \caption{Original production scan at $\omega=3.7\MeV$, $E_2$. Solid curves show exact fixed-time responses; dashed curves use archived single-peak fits. Error bars are binomial standard errors with pooled central queries; displayed IBM center errors are finite-shot bootstrap uncertainties. No refitting was performed.}
 \label{fig:scans}
\end{figure*}


For each of 2,000 parametric replicas,
$k_i^{(b)}\sim\mathrm{Binomial}(N_i,\widehat p_i)$,
where $\widehat p_i=P_{\rm fit}(x_i)$ is evaluated at the
original maximum-likelihood fit.
Approximate weighted-least-squares (WLS) refitting uses
fixed weights
\begin{equation}
 w_i=N_i/\max[\widehat p_i(1-\widehat p_i),10^{-6}]
 \label{eq:Sweights}
\end{equation}
and profiles the amplitude on 1001 centers $c\in\Eref\pm25\keV$:
\begin{equation}
 K_b(c)=\frac{\sum_iw_i(k_i^{(b)}/N_i)f_i(c)}
                  {\sum_iw_if_i(c)^2},\quad f_i(c)=f_{\mathbf t}(x_i-c).
 \label{eq:Sprofile}
\end{equation}
After clipping $K_b$ to its bounds, the minimum weighted residual selects
$c$; parabolic interpolation at an interior minimum refines the
$0.05\keV$ grid within one step. The reported center remains the original
binomial MLE, not the bootstrap mean. 

Independent profile-binomial refits of the two simulator fits
with optimizer termination warnings reproduce their centers within
$5\times10^{-7}\keV$, leaving the rounded tabulated results unchanged.
For FG--R6 in the additional simulator study, exact-MLE checks of
12 bootstrap replicas per target give absolute differences from the
profile-WLS centers below $0.041\keV$.
These checks concern the additional simulations, not a
replica-by-replica validation of the original QPU analysis.

\section{Modified effective range expansion (MERE)
reconstruction and uncertainty definitions}
\label{sec:mere}
\begin{figure*}[!tbp]
 \centering
 \suppgraphic{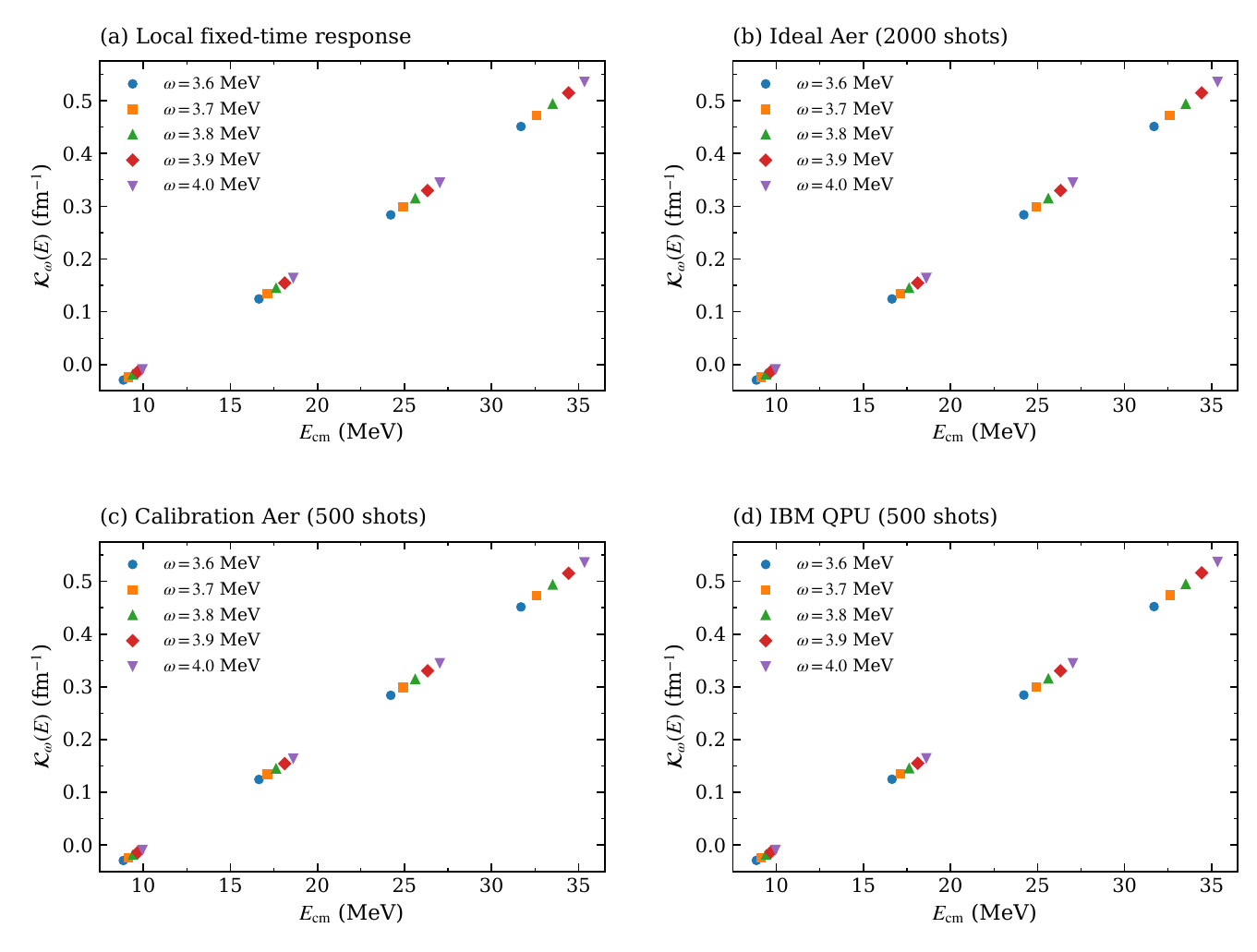}{0.98\textwidth}{66mm}
 \caption{Original FG--R6 finite-confinement inputs from four data sources, each with four levels at five trap strengths. These enter the two-stage MERE fit, not independent continuum measurements.}
 \label{fig:pcotinputs}
\end{figure*}

\begin{figure*}[!tbp]
 \centering
 \suppgraphic{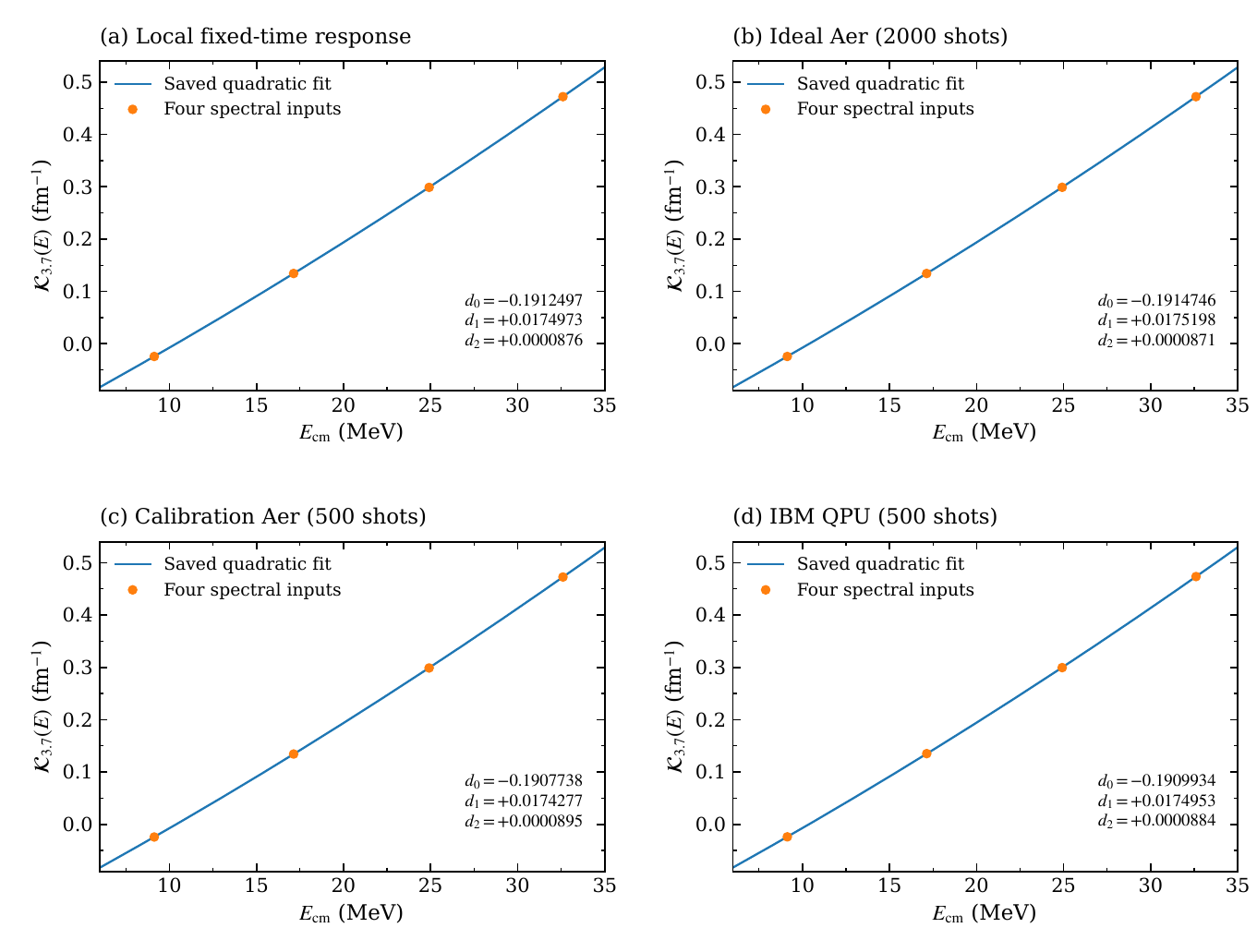}{0.98\textwidth}{50mm}
 \caption{First MERE stage: archived quadratic fits at $\omega=3.7\MeV$ for original FG--R6 sources. Coefficients use $E$ in MeV and $\mathcal K$ in $\mathrm{fm}^{-1}$; curves are least-squares fits, not arbitrary splines.}
 \label{fig:polyfits}
\end{figure*}

\begin{figure*}[!tbp]
 \centering
 \begin{minipage}[t]{0.485\textwidth}
  \centering
  \suppgraphic{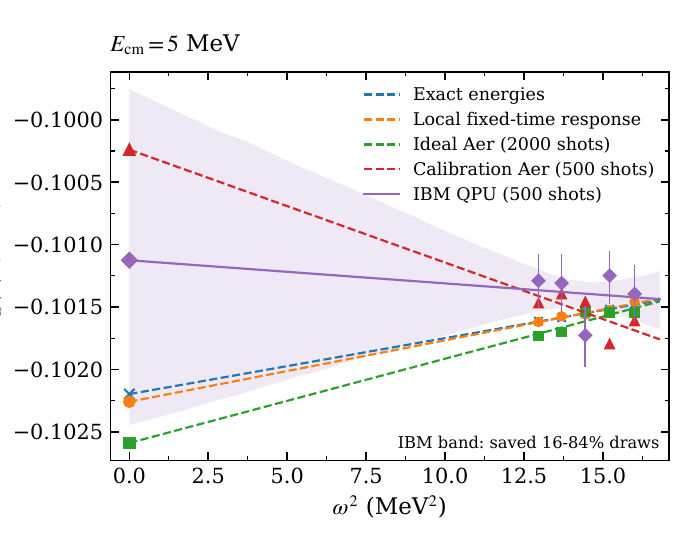}{\linewidth}{49mm}
  \par\smallskip (a) $E=5\MeV$
 \end{minipage}\hfill
 \begin{minipage}[t]{0.485\textwidth}
  \centering
  \suppgraphic{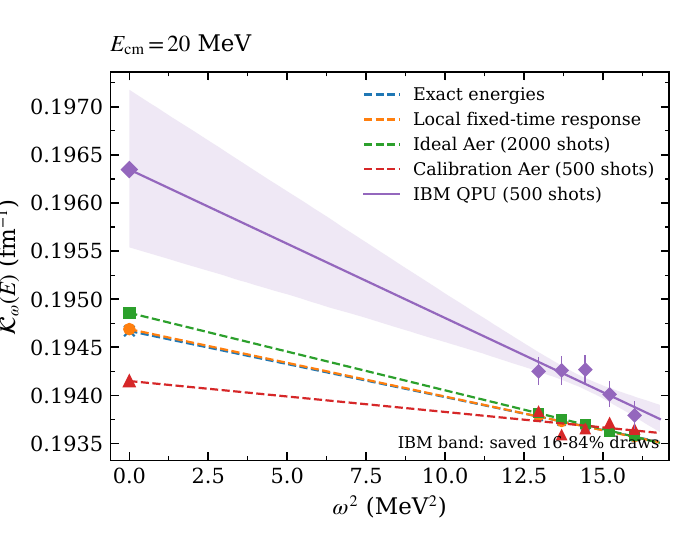}{\linewidth}{49mm}
  \par\smallskip (b) $E=20\MeV$
 \end{minipage}
 \caption{Second MERE stage: fixed-$E$ fits versus $\omega^2$ yield $K_0(E)$ at zero trap. The five points are energy-polynomial evaluations, not extra QPU measurements. The energy step extrapolates at 5.0 MeV and interpolates at 20.0 MeV; trap removal extrapolates in both. IBM intervals and shading are pointwise 16--84\% ranges from the original 1,500 archived MERE draws.
 Symbols at $\omega^2=0$ denote the fitted intercepts $K_0(E)$,
not additional spectral inputs.}
 \label{fig:trapremoval}
\end{figure*}

At each exact or fitted trapped energy, the finite-confinement input is
\cite{Luu2010,Zhang2020,Wang2024}
\begin{equation}
 \mathcal K_\omega(E_n)=-\frac{\sqrt{4\mu\omega}}{\hc}
 \frac{\Gamma(3/4-E_n/(2\omega))}{\Gamma(1/4-E_n/(2\omega))}.
 \label{eq:SKomega}
\end{equation}
This is in $\mathrm{fm}^{-1}$, as is $p$ in Eq.~\eqref{eq:Smomenta}.
Writing $\mathcal K_\omega=p\cot\delta_{0,\omega}$ does not make it a
free-space observable: the trap perturbs the interaction region.
Four points at each $\omega$ give an unweighted quadratic fit,
\begin{equation}
 \mathcal K_\omega(E)\simeq d_0(\omega)+d_1(\omega)E+d_2(\omega)E^2.
 \label{eq:Senergyfit}
\end{equation}
Evaluate all five polynomials at the \emph{same} $E$, then fit unweighted:
\begin{equation}
 \begin{aligned}
 \mathcal K_\omega(E)&\simeq K_0(E)+K_2(E)\omega^2,\\
 \delta_0^{\rm MERE}(E)&=\pi/2-\arctan[K_0(E)/p(E)].
 \end{aligned}\label{eq:Strapfit}
\end{equation}
The principal arctangent gives $0<\delta_0<\pi$ for $p>0$, equivalent to
$\operatorname{atan2}(p,K_0)$. Variable rescaling affects numerical
conditioning, not the weights or model. Neither following one level
$E_n(\omega)$ to zero trap nor evaluating (\ref{eq:SKomega}) at arbitrary
$E$ replaces this fixed-energy procedure (Figs.~\ref{fig:pcotinputs}--\ref{fig:trapremoval}).



Each of the 1,500 MERE replicas independently draws one stored
center for each of the twenty targets, reevaluates 
Eq.~(\ref{eq:SKomega}) at the sampled energies, and repeats both fits.
Pointwise 16--84\%
quantiles and standard deviations are conditional finite-shot errors;
cross-target hardware covariance, batch drift, and fit-form uncertainty
are excluded. Different energies on a curve share those twenty inputs
and are correlated, not independent measurements.

On $E=0.1,0.2,\ldots,30.0\MeV$, all five energy fits extrapolate below
$8.855470\MeV$ (88 points); support is mixed until $9.956592\MeV$
(11 points); common interpolation covers $10.0$--$30.0\MeV$ (201 points).
All interval metrics restrict the same curves without refitting, and
trap removal always extrapolates. Exact-energy MERE differs from the
analytical reference by at most $0.1411^\circ$, $0.0502^\circ$, and
$0.0372^\circ$ in these three regions; the maximum above the lowest
input is $0.0502^\circ$ and the full-grid relative maximum is $0.1128\%$.
These measure the fixed reference and primary prescription, not total
theoretical or QPU uncertainty.

\begin{figure*}[!tbp]
 \centering
 \suppgraphic{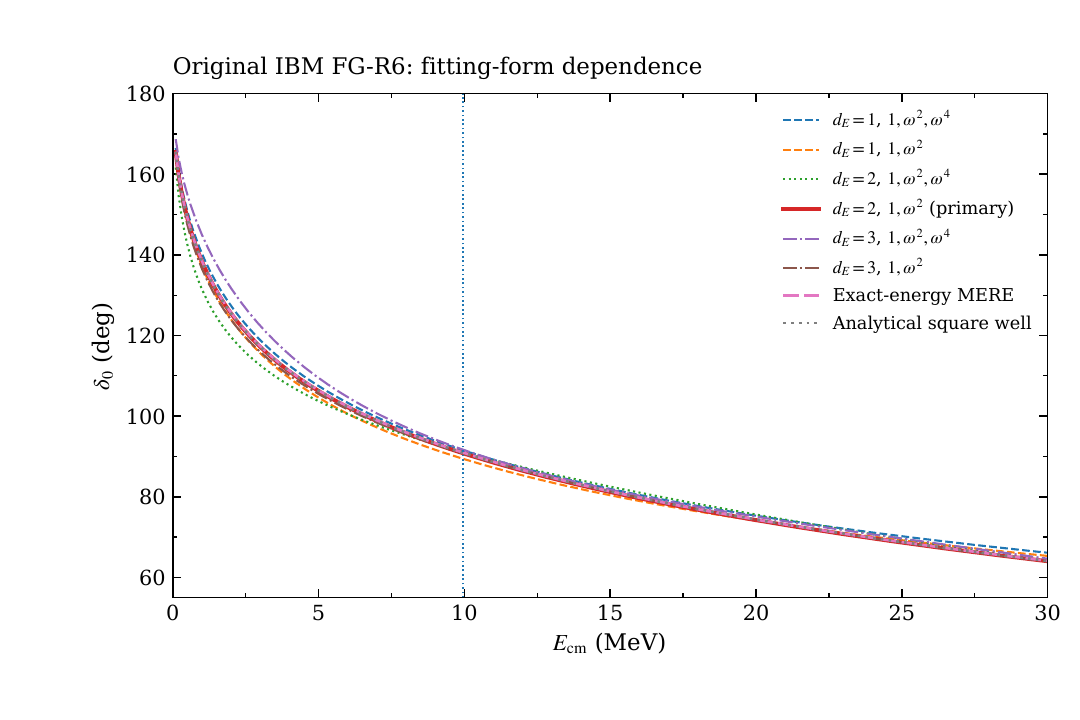}{0.94\textwidth}{62mm}
 \caption{Original IBM FG--R6 fit-form sensitivity: all six energy/trap variants are retained and the primary fit is identified. This envelope is not the finite-shot band of the main-text phase plot.
 Legend labels specify the energy-polynomial degree $d_E$
and the basis functions retained in the trap fit.
The vertical dotted line at $E_{\rm cm}=9.9566\MeV$
marks the onset of common energy interpolation;
trap removal remains an extrapolation throughout.
 }
 \label{fig:prodmodels}
\end{figure*}

Six variants combine energy degrees $1,2,3$ with trap powers
$\{0,2\}$ or $\{0,2,4\}$. Using a common phase convention,
\begin{equation}
 \begin{aligned}
 {}[\delta_{\min},\delta_{\max}](E)&=[\min_v\delta_v(E),\max_v\delta_v(E)],\\
 h_{\rm model}(\mathcal I)&=\max_{E\in\mathcal I}
       [\delta_{\max}(E)-\delta_{\min}(E)]/2.
 \end{aligned}\label{eq:Senvelope}
\end{equation}
All six are retained (Fig.~\ref{fig:prodmodels}). This is a deterministic
sensitivity diagnostic, not a confidence/credible interval or calibrated
truncation error, and is not added in quadrature to shot errors. A cubic
through four energies has no residual degree of freedom and can amplify
extrapolation sensitivity.

\section{Original production performance and hardware inputs}
\label{sec:production}
\begin{table*}[!tbp]
  \centering
  \caption{Original production summary. Energy entries are in keV, phase entries in degrees. $L_{\rm ref}=\max_E|\delta_0-\delta_0^{\rm ref}|$ uses the full 300-point grid; $h_{\rm model}$ is defined in Eq.~\eqref{eq:Senvelope}. Dashes mark quantities inapplicable to deterministic probabilities. Shots are per circuit.}
  \label{tab:fullperf}
  \footnotesize
  \setlength{\tabcolsep}{4pt}
  \begin{tabular}{@{}llrrrrrrr@{}}
    \toprule
    Source & Method & Shots & RMS $\Delta E_c$ & Max. $|\Delta E_c|$ & Median $\sigma_E$ & Median $K$ & $L_{\rm ref}$ & $h_{\rm model}$ \\
    \midrule
    Local formula & Direct R10 & -- & 0.0237 & 0.0788 & -- & 0.8930 & 0.00345 & 1.054 \\
    Local formula & FG--R6 & -- & 0.0334 & 0.0853 & -- & 0.8947 & 0.01742 & 1.072 \\
    Local sampled & Direct R10 & 2000 & 0.2154 & 0.6043 & 0.2023 & 0.8908 & 0.15175 & 3.723 \\
    Local sampled & FG--R6 & 2000 & 0.1807 & 0.4299 & 0.1993 & 0.8960 & 0.02478 & 3.310 \\
    Ideal Aer & Direct R10 & 2000 & 0.2188 & 0.4000 & 0.2056 & 0.8923 & 0.09559 & 1.437 \\
    Ideal Aer & FG--R6 & 2000 & 0.1316 & 0.2934 & 0.2015 & 0.8987 & 0.11817 & 4.524 \\
    Calibration Aer & Direct R10 & 500 & 0.4913 & 0.9110 & 0.5000 & 0.7039 & 0.52103 & 8.805 \\
    Calibration Aer & FG--R6 & 500 & 0.4930 & 1.2165 & 0.4576 & 0.7763 & 0.53408 & 11.390 \\
    IBM QPU & Direct R10 & 500 & 0.7934 & 1.6380 & 0.6099 & 0.5371 & 0.84226 & 20.119 \\
    IBM QPU & FG--R6 & 500 & 1.6476 & 2.5517 & 0.5342 & 0.6520 & 0.16347 & 6.753 \\
    \bottomrule
  \end{tabular}
\end{table*}

Table~\ref{tab:fullperf} keeps the original sources separate. Even noiseless
full-spectrum data can yield nonzero single-peak bias because
Eq.~\eqref{eq:Speak} omits other states. Local sampling isolates shot
noise; IBM F6 has larger amplitude and better median shot precision but
larger energy RMS than D10, not reproduced by calibration Aer.
The signs in Table~\ref{tab:qpuinputs} establish neither covariance nor
error cancellation. These twenty fitted energies supply each IBM curve;
none is replaced by an exact eigenvalue or sampled-subspace result.

\begin{table}[!htbp]
  \centering
  \caption{Original IBM spectral inputs: reference energies in MeV; signed offsets and finite-shot errors in keV. Thus $E_c=E_{\rm ref}+10^{-3}\Delta E_c$. Reproduction uses full-precision files, not these rounded values.}
  \label{tab:qpuinputs}
  \footnotesize
  \setlength{\tabcolsep}{2pt}
  \begin{tabular}{@{}ccrrr@{}}
    \toprule
    $\omega$ & $n$ & $E_{\rm ref}$ & R10: $\Delta E_c\pm\sigma_E$ & F6: $\Delta E_c\pm\sigma_E$ \\
    \midrule
    3.6 & $E_1$ & 8.855470 & $+0.153\pm0.576$ & $+1.838\pm0.506$ \\
    3.6 & $E_2$ & 16.643000 & $+1.142\pm0.591$ & $+1.371\pm0.521$ \\
    3.6 & $E_3$ & 24.213795 & $+0.450\pm0.651$ & $+1.682\pm0.539$ \\
    3.6 & $E_4$ & 31.691304 & $+0.254\pm0.593$ & $+1.766\pm0.539$ \\
    \addlinespace[2pt]
    3.7 & $E_1$ & 9.129551 & $+0.595\pm0.615$ & $+1.271\pm0.506$ \\
    3.7 & $E_2$ & 17.134039 & $-1.207\pm0.643$ & $+2.552\pm0.530$ \\
    3.7 & $E_3$ & 24.916082 & $+0.306\pm0.612$ & $+1.276\pm0.535$ \\
    3.7 & $E_4$ & 32.602105 & $+0.079\pm0.613$ & $+2.433\pm0.546$ \\
    \addlinespace[2pt]
    3.8 & $E_1$ & 9.404443 & $-1.055\pm0.589$ & $+0.987\pm0.498$ \\
    3.8 & $E_2$ & 17.625943 & $-0.358\pm0.599$ & $+1.629\pm0.534$ \\
    3.8 & $E_3$ & 25.619273 & $+0.704\pm0.656$ & $+2.035\pm0.575$ \\
    3.8 & $E_4$ & 33.513827 & $-0.975\pm0.618$ & $+0.980\pm0.552$ \\
    \addlinespace[2pt]
    3.9 & $E_1$ & 9.680129 & $+0.090\pm0.605$ & $+1.209\pm0.487$ \\
    3.9 & $E_2$ & 18.118693 & $-1.638\pm0.643$ & $+1.716\pm0.497$ \\
    3.9 & $E_3$ & 26.323347 & $-0.653\pm0.665$ & $+1.311\pm0.503$ \\
    3.9 & $E_4$ & 34.426445 & $-0.074\pm0.620$ & $+2.178\pm0.543$ \\
    \addlinespace[2pt]
    4.0 & $E_1$ & 9.956592 & $+0.749\pm0.584$ & $-0.068\pm0.498$ \\
    4.0 & $E_2$ & 18.612272 & $+0.856\pm0.586$ & $+1.648\pm0.562$ \\
    4.0 & $E_3$ & 27.028285 & $-1.066\pm0.566$ & $+0.624\pm0.571$ \\
    4.0 & $E_4$ & 35.339939 & $+0.920\pm0.608$ & $+2.168\pm0.537$ \\
    \bottomrule
  \end{tabular}
\end{table}
\begin{table*}[!tbp]
  \centering
  \caption{Original IBM phase metrics (degrees), as in the main text. $\Delta\delta_0=\delta_0^{\rm QPU}-\delta_0^{\rm ref}$. The full and common-interpolation subsets contain 300 and 201 points from the same curves, without refitting.}
  \label{tab:qpuregions}
  \footnotesize
  \setlength{\tabcolsep}{4pt}
  \begin{tabular}{@{}llrrrrr@{}}
    \toprule
    $E$ (MeV) & Method & Max. $|\Delta\delta_0|$ & RMS $\Delta\delta_0$ & Max. $|\delta_0^{\rm QPU}-\delta_0^{\rm an}|$ & Max. $\sigma_\delta$ & $h_{\rm model}$ \\
    \midrule
    $0.1$--$30.0$ & Direct R10 & 0.842 & 0.305 & 0.981 & 0.429 & 20.12 \\
    $0.1$--$30.0$ & FG--R6 & 0.163 & 0.137 & 0.156 & 0.401 & 6.75 \\
    $10.0$--$30.0$ & Direct R10 & 0.137 & 0.102 & 0.160 & 0.074 & 3.19 \\
    $10.0$--$30.0$ & FG--R6 & 0.159 & 0.129 & 0.156 & 0.065 & 1.12 \\
    \bottomrule
  \end{tabular}
\end{table*}

The full-grid exact-MERE residual maxima, $0.842^\circ$ (R10) and
$0.163^\circ$ (F6), occur at 1.2 and 6.1 MeV, below all inputs.
On the common-interpolation subset, R10 has slightly smaller maximum
and RMS reference residuals, while maxima against the analytical
solution are similar, $0.160^\circ/0.156^\circ$
(Table~\ref{tab:qpuregions}). Model half-ranges are $20.12^\circ/6.75^\circ$
on the full grid and $3.19^\circ/1.12^\circ$ on the subset. Thus the
full-grid central improvement is not uniform or a total-error reduction.

\begin{figure*}[!tbp]
 \centering
 \suppgraphic{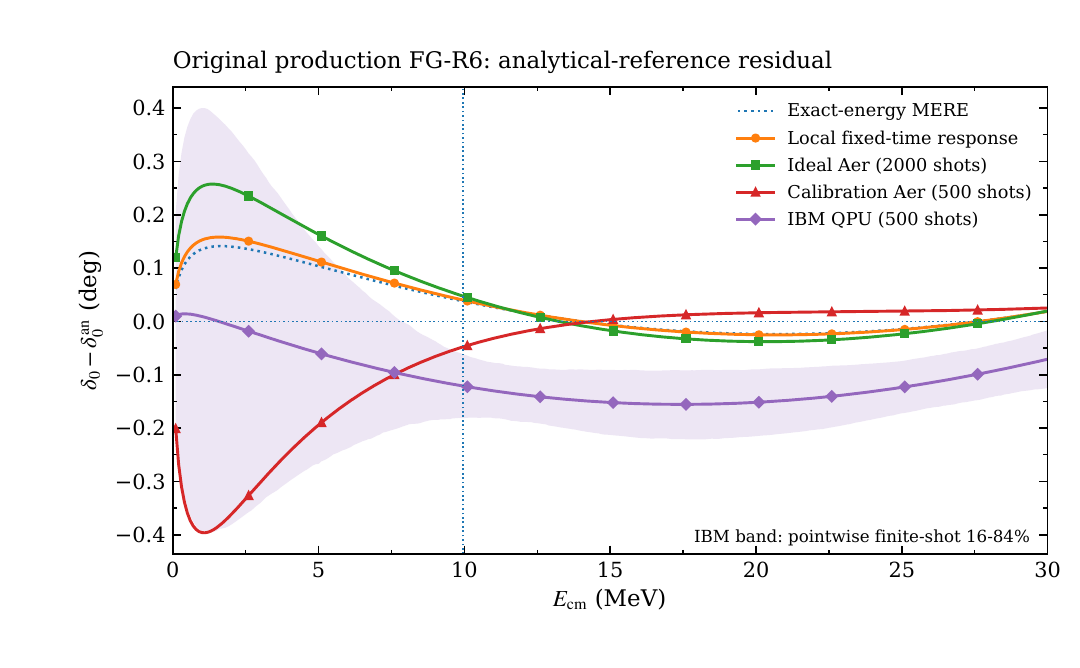}{0.85\textwidth}{56mm}
 \caption{Original production FG--R6 residuals relative to the analytical square-well phase. The main-text residual instead uses exact-energy MERE. These sources are distinct from the retrospective synthetic-noise study.
 The vertical dotted line marks the onset of common energy
interpolation at $E_{\rm cm}=9.9566\MeV$.
 }
 \label{fig:analyticresidual}
\end{figure*}

Full-grid analytical residual maxima (R10/F6) are
$0.138/0.158^\circ$ for local response, $0.237/0.258^\circ$ for ideal Aer,
$0.660/0.396^\circ$ for calibration Aer, and $0.981/0.156^\circ$ for IBM
(Fig.~\ref{fig:analyticresidual}). They use a different reference from
$L_{\rm ref}$; maxima cannot be added or subtracted to convert between them.

\section{Retrospective justification of the six-cycle sequence}
\label{sec:cycle}
The unchanged production R6 is compared with fixed-$R=4,5,6,7,8$
designs using the same loss, bounds, and one optimizer seed. The retuned
R4 ($t_{\max}=31.7717529157\,\mathrm{MeV}^{-1}$, $\alpha=1.015$)
was separately selected using the known four-level spectrum, unlike the
isolated-response loss. This adds no hardware data.
Deterministic probabilities at seven queries are fitted by profile least
squares, not finite-shot binomial MLE. Anchor shifts of
$-10,-5,0,5,10\keV$ leave the Hamiltonian and true target fixed;
errors are measured from that target, not the shifted anchor.

\begin{table*}[!tbp]
  \centering
  \caption{Retrospective cycle tests with the original loss. $D_{\max}$ is the matching-grid response difference; $B_0$ and $B_{\rm shift}$ are maximum four-level center errors (keV) at the nominal and shifted anchors. $t_{\max}$ is in $\mathrm{MeV}^{-1}$. Retuned R4 uses the known spectrum.}
  \label{tab:cycles}
  \small
  \setlength{\tabcolsep}{4pt}
  \begin{tabular}{@{}lrrrrr@{}}
    \toprule
    Sequence & $t_{\max}$ & $\alpha$ & $D_{\max}$ & $B_0$ (keV) & $B_{\rm shift}$ (keV) \\
    \midrule
    R10 reference & -- & -- & 0 & 0.07882 & 0.08623 \\
    R4, original loss & 31.542746 & 1.010000 & $7.01\times10^{-4}$ & 0.12303 & 0.13505 \\
    R4, retuned & 31.771753 & 1.015000 & $7.09\times10^{-4}$ & 0.08986 & 0.09577 \\
    R5, original loss & 35.000916 & 1.139906 & $1.79\times10^{-4}$ & 0.31481 & 0.36376 \\
    R6, production & 36.505394 & 1.199231 & $8.51\times10^{-6}$ & 0.08533 & 0.09255 \\
    R7, original loss & 37.232115 & 1.228648 & $1.90\times10^{-4}$ & 0.07222 & 0.08738 \\
    R8, original loss & 37.712886 & 1.246506 & $4.00\times10^{-4}$ & 0.03128 & 0.03996 \\
    \bottomrule
  \end{tabular}
\end{table*}

Under the original loss, returned R4/R5 fail the inherited $0.10\keV$
center limit, whereas R6 passes even at shifted anchors and matches the
local R10 response much more closely (Table~\ref{tab:cycles}). Retuned R4
also passes: no minimum-cycle or global-optimality claim follows from
these single-seed tests. R8 improves several ideal measures at greater
cycle cost. Tolerances were not tightened after inspecting shorter candidates.

Local resolution is retained:
\begin{equation}
 \log f_{\mathbf t}(x)=-\frac{x^2}{4}\sum_jt_j^2
             -\frac{x^4}{96}\sum_jt_j^4+O(x^6).
 \label{eq:Smoments}
\end{equation}
R10/R6 second moments are $3879.8622/3879.7596\,\mathrm{MeV}^{-2}$;
the median isolated-peak Fisher lower bounds for the actual query
allocation and overlaps are $0.402008/0.402010\keV$. These describe
ideal local precision, not hardware errors.

\subsection{Declared off-target domains}
\begin{figure*}[!tbp]
 \centering
 \begin{minipage}[t]{0.485\textwidth}
  \centering
  \suppgraphic{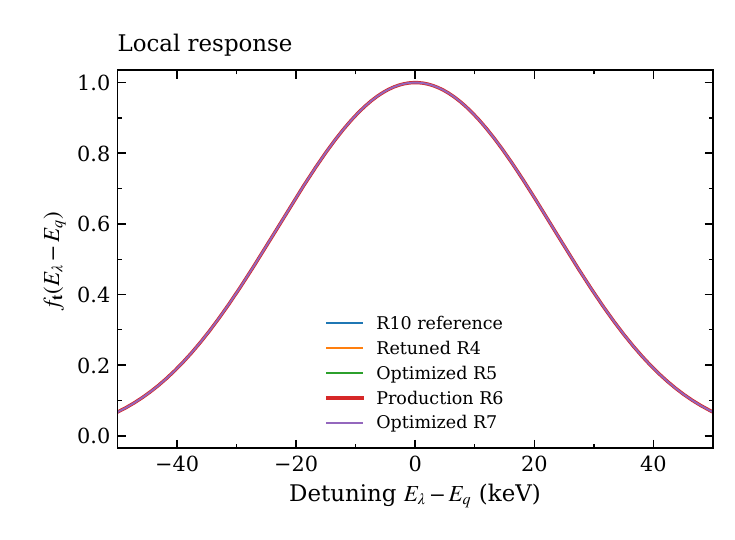}{\linewidth}{47mm}
  \par\smallskip (a) Local response
 \end{minipage}\hfill
 \begin{minipage}[t]{0.485\textwidth}
  \centering
  \suppgraphic{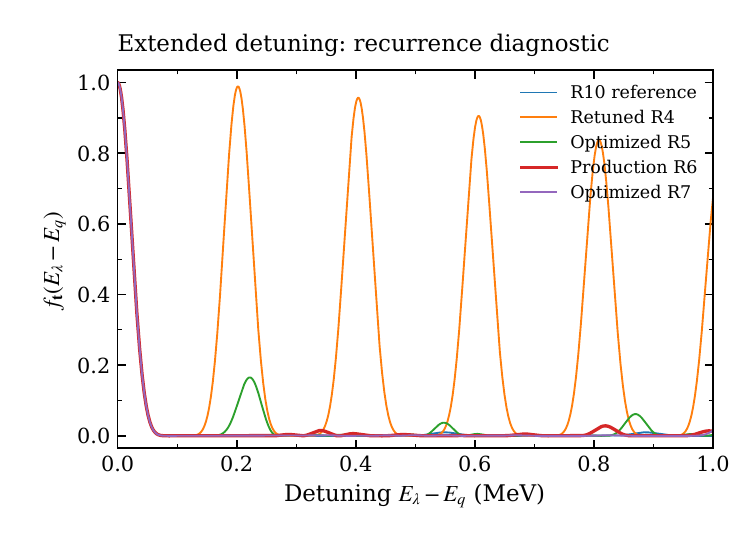}{\linewidth}{47mm}
  \par\smallskip (b) Extended detuning range
 \end{minipage}
 \caption{Retrospective filter comparison. Local agreement does not ensure off-target agreement. The 0--1 MeV panel shows recurrences, not actual interlevel spacings or a suppression certificate; Table~\ref{tab:offtarget} uses the declared spectral intervals.}
 \label{fig:filters}
\end{figure*}

Each non-target gap $|E_m-E_n|$ is broadened by the $\pm40\keV$ query
range plus 10 keV padding per side; overlapping intervals are merged
in \texttt{detuning\_domains.json}. Positive intervals span approximately
7.4275--25.4333 MeV with gaps; evenness of $f$ covers negative detuning.
With grid spacing at most $h=10^{-4}\MeV$,
\begin{align}
 |f_{\mathbf t}'(x)|&\leq\tfrac12\sum_j|t_j|,\label{eq:Sderivative}\\
 U_{\rm off}&=\min[1,\max_{\rm grid}f_{\mathbf t}(x)
                      +\tfrac h4\sum_j|t_j|].\label{eq:Soffbound}
\end{align}
Bounds add when comparing two filters. These are enclosures only on
the declared domains; no mandatory off-target tolerance was enabled.
The weighted non-target contribution at $E_q=E_n$ is
\begin{equation}
 \eta_{\rm off}=\max_{n,\omega}\sum_{m\ne n}|c_m|^2
                                  f_{\mathbf t}(E_m-E_n).
 \label{eq:Seta}
\end{equation}
It differs from $U_{\rm off}$ and from conditional-state infidelity
(Table~\ref{tab:offtarget}). Figure~\ref{fig:filters} shows local response
and recurrences. R6 is a tested accuracy--resource compromise for this
spectrum/window; other overlaps, Hamiltonians, or required scattering
accuracy can require different times or $R$. Factorization does not
single out six cycles.

\begin{table}[!htbp]
  \centering
  \caption{Ideal off-target diagnostics on the declared domains: transmission upper bound $U_{\rm off}$ and weighted non-target probability $\eta_{\rm off}$ at central queries. Neither is a conditional-state infidelity; no mandatory off-target tolerance was imposed.}
  \label{tab:offtarget}
  \small
  \setlength{\tabcolsep}{4pt}
  \begin{tabular}{@{}lrr@{}}
    \toprule
    Sequence & $U_{\rm off}$ & $\eta_{\rm off}$ \\
    \midrule
    R10 reference & 0.05675 & 0.001265 \\
    R4, original loss & 0.25514 & 0.008530 \\
    R4, retuned & 0.54326 & 0.007250 \\
    R5, original loss & 0.38644 & 0.014574 \\
    R6, production & 0.31651 & 0.003184 \\
    R7, original loss & 0.42230 & 0.021778 \\
    R8, original loss & 0.04347 & 0.000792 \\
    \bottomrule
  \end{tabular}
\end{table}

\section{Additional simulator comparisons and statistical checks}
\label{sec:simulators}
All sampled retrospective sources use 500 shots per occurrence on the
same targets/queries, unlike the original ideal/local-sampled controls.
The methods are D10/F10 (direct/factorized R10), D6/F6 (direct/factorized
with production R6 times), F4C (spectrum-informed R4), F5O/F7O (optimized
R5/R7), and F6S (signed R6). R4O/R8O enter only classical tests.
F6S multiplies the R6 times by $(+,-,-,+,-,+)$; cosine squares preserve
ideal all-zero probabilities, not necessarily the full accepted-state
operator. It is neither an IBM modification nor established mitigation.
Exploratory SGC--R6 files describe a different spectrum-guided sequence.

\subsection{Independent density and synthetic-noise models}
The complete three-qubit density-matrix propagation includes controlled
evolution, phase, measurement and reset, agreeing with the fixed-time
formula. Local sampling draws binomial ideal counts. Cycle-level noise
uses $\rho\mapsto(1-\epsilon)\rho+\epsilon\operatorname{tr}(\rho)I/8$
with $\epsilon=0.01$ per cycle, ancilla assignment error 0.015, and reset
excitation 0.005. The accepted branch is the \emph{recorded} all-zero
branch, including assignment errors, not an ad hoc attenuation.

Offline Aer compilation uses level 3, seed 20260912, and
$\{R_z,\mathrm{SX},X,\mathrm{CX}\}$ without IBM topology. Synthetic noise
uses depolarizing parameters $2\times10^{-4}$ on SX/X and
$3\times10^{-3}$ on CX, symmetric readout error 0.01, and X after reset
with probability 0.005; $R_z$ has no added stochastic error. These are
assumed dimensionless parameters, not measured infidelities or IBM
calibration. Recorded versions are Python 3.11.15, NumPy 2.4.6,
SciPy 1.17.1, Qiskit 2.5.1, Aer 0.17.2, Runtime 0.48.0; base seed
20260912 and task/source seeds are archived. Energy/MERE replica counts
remain 2000/1500.

\begin{table}[!htbp]
  \centering
  \caption{Retrospective simulator results, separate from production (Table~\ref{tab:fullperf}). Energy entries are in keV. All sampled sources use 500 shots per occurrence; deterministic quantities have no shot error. $L_{\rm ref}$ and $h_{\rm model}$ are full-grid degrees. 
  The flagged simulator fits were independently checked without
  changing the rounded entries (Sec.~S5).
  }
  \label{tab:studysummary}
  \footnotesize
  \setlength{\tabcolsep}{2pt}
  \begin{tabular}{@{}lrrrrrr@{}}
    \toprule
    Method & $R$ & RMS $\Delta E_c$ & Med. $\sigma_E$ & Med. $K$ & $L_{\rm ref}$ & $h_{\rm model}$ \\
    \midrule
    \multicolumn{7}{@{}l}{\textit{Deterministic density matrix}} \\
    D10 & 10 & 0.0237 & -- & 0.8930 & 0.00345 & 1.054 \\
    F10 & 10 & 0.0237 & -- & 0.8930 & 0.00345 & 1.054 \\
    D6 & 6 & 0.0334 & -- & 0.8947 & 0.01742 & 1.072 \\
    F6 & 6 & 0.0334 & -- & 0.8947 & 0.01742 & 1.072 \\
    F4C & 4 & 0.0449 & -- & 0.8954 & 0.07909 & 1.054 \\
    F5O & 5 & 0.1389 & -- & 0.8968 & 0.07488 & 3.997 \\
    F7O & 7 & 0.0283 & -- & 0.8966 & 0.03093 & 1.769 \\
    F6S & 6 & 0.0334 & -- & 0.8947 & 0.01742 & 1.072 \\
    \midrule
    \multicolumn{7}{@{}l}{\textit{Local binomial sampling}} \\
    D10 & 10 & 0.3662 & 0.4060 & 0.8907 & 0.77757 & 3.914 \\
    F10 & 10 & 0.3083 & 0.4015 & 0.8909 & 0.16326 & 6.523 \\
    D6 & 6 & 0.4544 & 0.4048 & 0.8913 & 0.27730 & 5.784 \\
    F6 & 6 & 0.3580 & 0.4002 & 0.8989 & 0.11661 & 9.036 \\
    F4C & 4 & 0.4060 & 0.4024 & 0.8917 & 0.15754 & 6.641 \\
    F5O & 5 & 0.4265 & 0.3993 & 0.8978 & 0.21506 & 5.761 \\
    F7O & 7 & 0.3319 & 0.4031 & 0.8939 & 0.08197 & 3.321 \\
    F6S & 6 & 0.4002 & 0.4061 & 0.8876 & 0.08009 & 25.788 \\
    \midrule
    \multicolumn{7}{@{}l}{\textit{Cycle-level assumed noise}} \\
    D10 & 10 & 0.4797 & 0.5153 & 0.6829 & 0.62365 & 6.449 \\
    F10 & 10 & 0.4666 & 0.5106 & 0.6912 & 0.57843 & 6.271 \\
    D6 & 6 & 0.6373 & 0.4637 & 0.7642 & 0.14399 & 65.307 \\
    F6 & 6 & 0.4069 & 0.4659 & 0.7674 & 0.15785 & 8.031 \\
    F4C & 4 & 0.3612 & 0.4455 & 0.8094 & 0.33733 & 6.056 \\
    F5O & 5 & 0.4688 & 0.4555 & 0.7918 & 0.76957 & 7.910 \\
    F7O & 7 & 0.4835 & 0.4751 & 0.7437 & 0.27558 & 41.057 \\
    F6S & 6 & 0.3784 & 0.4603 & 0.7659 & 0.05048 & 11.036 \\
    \midrule
    \multicolumn{7}{@{}l}{\textit{Ideal Aer (500 shots)}} \\
    D10 & 10 & 0.4237 & 0.4013 & 0.8930 & 0.27816 & 1.473 \\
    F10 & 10 & 0.4228 & 0.3995 & 0.9009 & 0.38658 & 3.190 \\
    D6 & 6 & 0.4723 & 0.3996 & 0.8901 & 0.31652 & 8.923 \\
    F6 & 6 & 0.4770 & 0.4033 & 0.8885 & 0.06947 & 12.480 \\
    F4C & 4 & 0.2619 & 0.4022 & 0.8957 & 0.24655 & 3.329 \\
    F5O & 5 & 0.3618 & 0.4007 & 0.9005 & 0.32664 & 22.162 \\
    F7O & 7 & 0.3543 & 0.3991 & 0.8985 & 0.27549 & 17.752 \\
    F6S & 6 & 0.4329 & 0.4010 & 0.8949 & 0.19736 & 11.633 \\
    \midrule
    \multicolumn{7}{@{}l}{\textit{Synthetic noisy Aer (500 shots)}} \\
    D10 & 10 & 0.3460 & 0.5159 & 0.6810 & 0.21115 & 12.370 \\
    F10 & 10 & 0.6144 & 0.5119 & 0.6775 & 0.52271 & 11.054 \\
    D6 & 6 & 0.4961 & 0.4636 & 0.7645 & 0.60839 & 70.771 \\
    F6 & 6 & 0.5239 & 0.4677 & 0.7591 & 0.39983 & 64.448 \\
    F4C & 4 & 0.4232 & 0.4398 & 0.8098 & 0.22516 & 9.192 \\
    F5O & 5 & 0.5547 & 0.4556 & 0.7961 & 0.61483 & 46.785 \\
    F7O & 7 & 0.3861 & 0.4803 & 0.7425 & 0.29006 & 2.596 \\
    F6S & 6 & 0.3735 & 0.4646 & 0.7642 & 0.09890 & 9.670 \\
    \bottomrule
  \end{tabular}
\end{table}
\begin{table}[!htbp]
  \centering
  \caption{Offline compilation in the $\{R_z,\mathrm{SX},X,\mathrm{CX}\}$ basis, without IBM topology. Gate--shots sum two-qubit gates times 500 shots over 160 entries; this is not execution time.}
  \label{tab:offlinecost}
  \small
  \setlength{\tabcolsep}{4pt}
  \begin{tabular}{@{}lrrr@{}}
    \toprule
    Method & Median depth & Mean 2Q gates & 2Q gate--shots \\
    \midrule
    D10 & 202.0 & 63.6 & 5\,088\,000 \\
    F10 & 208.5 & 63.6 & 5\,088\,000 \\
    D6 & 130.0 & 40.0 & 3\,200\,000 \\
    F6 & 128.5 & 40.0 & 3\,200\,000 \\
    F4C & 92.0 & 28.1 & 2\,248\,000 \\
    F5O & 110.5 & 33.0 & 2\,640\,000 \\
    F7O & 147.0 & 44.3 & 3\,544\,000 \\
    F6S & 129.0 & 40.0 & 3\,200\,000 \\
    \bottomrule
  \end{tabular}
\end{table}

F6 improves amplitude and shot precision in both assumed noise models,
not all phase errors (Table~\ref{tab:studysummary}); synthetic-Aer
$L_{\rm ref}$ is $0.39983^\circ$ versus $0.21115^\circ$ for D10.
Identical-time direct/factorized probabilities agree ideally, while shot
estimates differ. Six cycles lower offline gate--shot cost, but phase
factorization alone does not (Table~\ref{tab:offlinecost}); these costs
do not replace IBM topology-mapped resources.

\subsection{Independent shot realizations and method contrasts}
One hundred independent noiseless shot realizations show comparable
D10/F6 energy precision (Table~\ref{tab:mc}), not hardware repeatability.
For $L_m=\max_E|\delta_m-\delta_0^{\rm ref}|$ on the full grid,
independent-target bootstrap contrasts all span zero
(Table~\ref{tab:contrast}); central rankings are not universal accuracy
or reproducibility claims.

\begin{table}[!htbp]
  \centering
  \caption{One hundred independent noiseless shot realizations: mean twenty-level energy RMS (keV) and the distribution (degrees) of full-grid $L_{\rm ref}$. These are not hardware repeats.}
  \label{tab:mc}
  \footnotesize
  \setlength{\tabcolsep}{2.2pt}
  \begin{tabular}{@{}lrrl@{}}
    \toprule
    Method & Mean RMS$_E$ & Median $L_{\rm ref}$ & 16--84\% range \\
    \midrule
    D10 & 0.4056 & 0.2044 & [0.0868, 0.4059] \\
    F6 & 0.4027 & 0.1774 & [0.0718, 0.4061] \\
    \bottomrule
  \end{tabular}
\end{table}
\begin{table}[!htbp]
  \centering
  \caption{Conditional bootstrap contrast $L_{\rm D10}-L_{\rm F6}$ (degrees); positive favors F6. Quantiles exclude run-to-run hardware variability.}
  \label{tab:contrast}
  \small
  \setlength{\tabcolsep}{4pt}
  \begin{tabular}{@{}lrrr@{}}
    \toprule
    Source & Median & 2.5\% & 97.5\% \\
    \midrule
    Ideal Aer & 0.0776 & -0.4939 & 0.7046 \\
    Cycle-level noise & 0.3121 & -0.4883 & 1.1058 \\
    Synthetic noisy Aer & -0.0558 & -0.9031 & 0.6962 \\
    \bottomrule
  \end{tabular}
\end{table}

\subsection{High-order extrapolation sensitivity}
\label{sec:largespread}
\begin{figure*}[!tbp]
 \centering
 \suppgraphic{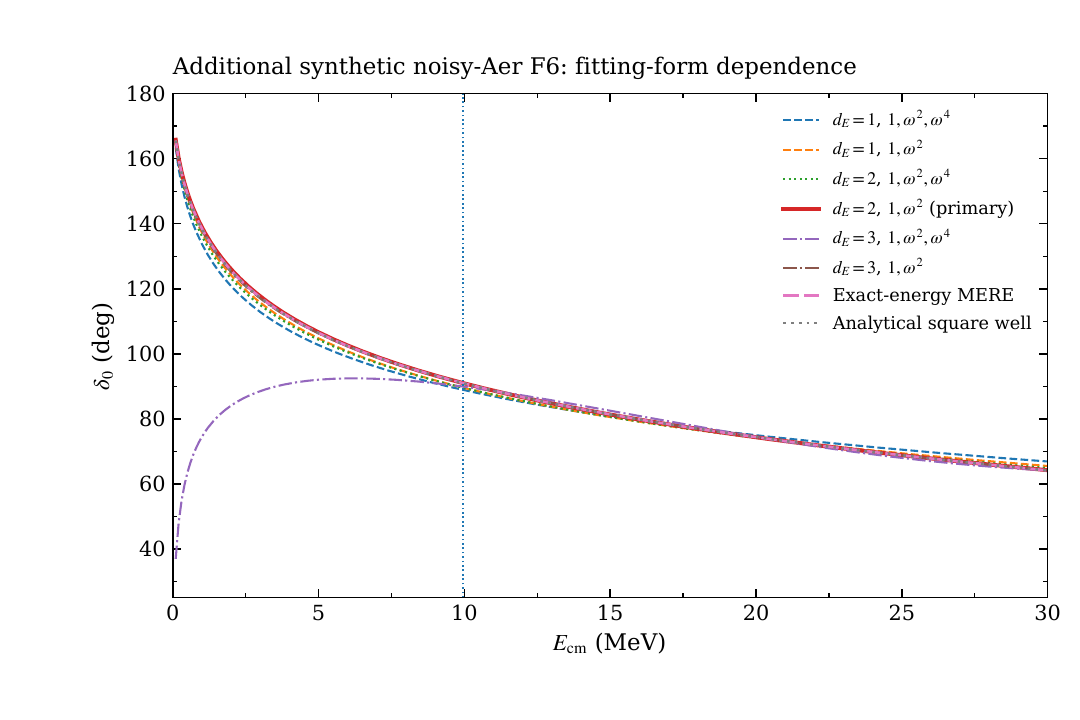}{0.94\textwidth}{62mm}
 \caption{Retrospective synthetic noisy-Aer F6 fit-form sensitivity, not IBM production. All six saved curves and both references are shown. The half-range reaches $64.4481^\circ$ at 0.1 MeV, separately from shot uncertainty and hardware repeatability.
 Legend labels specify the energy-polynomial degree $d_E$
 and the basis functions retained in the trap fit.
 The vertical dotted line marks the onset of common energy
 interpolation at $E_{\rm cm}=9.9566\MeV$.
 }
 \label{fig:syntheticmodels}
\end{figure*}

Synthetic-Aer F6 reaches $h_{\rm model}=64.4481^\circ$ at $E=0.1\MeV$
(Fig.~\ref{fig:syntheticmodels}): primary and cubic-energy/$\omega^4$
phases are $165.8046^\circ$ and $36.9084^\circ$, with intercepts
$-0.194127$ and $+0.065382\,\mathrm{fm}^{-1}$. This is a change in
$\mathcal K$, not merely a phase-branch plotting error; exact inputs
with the latter model give $165.5393^\circ$.
For $K_0=\sum_iw_i\mathcal K_{\omega_i}$, the trap weights are
\begin{eqnarray}
\begin{aligned}
 \mathbf{w}_{02}={}&(3.9543,\,2.1272,\,0.2501,\\
                 &\quad -1.6771,\,-3.6544),\\
 \mathbf{w}_{024}={}&(55.6542,\,-21.5562,\,-51.3039,\\
                  &\quad -29.5692,\,47.7752).
\end{aligned}
\end{eqnarray}
Their norms 6.03/96.70, combined with cubic energy extrapolation, explain
strong sensitivity to small spectral perturbations. On the common
interpolation subset, the synthetic F6 half-range remains $1.4694^\circ$
versus primary residual $0.06219^\circ$. The full spread is retained,
not removed, and is distinct from the original IBM $6.75^\circ$.

